\def\etal{{\it et al.\thinspace}}
\def\eg{{\it e.g., }}

\def\ie{{\it i.e., }}

\def\gcm3{{g cm${}^{-3}$}}
\def\h50{\hbox{$\rm\thinspace h_{50}$}}
\def\h50m1{\hbox{$\rm\thinspace h_{50}^{-1}$}}
\def\etal{{\it et al.\thinspace}}

\def\p3m{P${}^3$M}
\def\ap3m{AP${}^3$M}
\def\-{{\em{---}}}

\def\lsim{~\rlap{$<$}{\lower 1.0ex\hbox{$\sim$}}}
\def\gsim{~\rlap{$>$}{\lower 1.0ex\hbox{$\sim$}}}

\def\do{{\tt DO }}

\documentstyle[emulateapj,psfig,apjfonts]{article}

\received{}
\accepted{}
\lefthead{1 km Planetesimal Growth at 0.4 AU}
\righthead{Barnes \etal }

\begin{document}

\title{$N$-Body Simulations of Growth from 1 km Planetesimals at 0.4 AU}

\author{Rory Barnes\altaffilmark{1,2,3}, Thomas R. Quinn\altaffilmark{1}, 
Jack J. Lissauer\altaffilmark{4}, Derek C. Richardson\altaffilmark{5}}

\begin{center}E-mail:rory@astro.washington.edu
\end{center}

\altaffiltext{1}{Astronomy Department, University of Washington, Seattle, WA 98195}
\altaffiltext{2}{Lunar and Planetary Laboratory, University of Arizona, Tucson, AZ 85721}
\altaffiltext{3}{Virtual Planetary Laboratory}
\altaffiltext{4}{Space Science and Astrobiology Division, 245-3, NASA Ames Research Center, Moffett Field, CA 94035}
\altaffiltext{5}{Department of Astronomy, University of Maryland, College Park, MD 20742}

\begin{abstract}
We present $N$-body simulations of planetary accretion beginning with
1 km radius planetesimals in orbit about a 1 M$_{\odot}$ star at 0.4
AU. The initial disk of planetesimals contains too many bodies for any
current $N$-body code to integrate; therefore, we model a sample patch
of the disk. Although this greatly reduces the number of bodies, we
still track in excess of $10^5$ particles. We consider three initial velocity
distributions and monitor the growth of the planetesimals. The masses
of some particles increase by more than a factor of 100. Additionally,
the escape speed of the largest particle grows considerably faster
than the velocity dispersion of the particles, suggesting impending
runaway growth, although no particle grows large enough to detach
itself from the power law size-frequency distribution. These results are in general
agreement with previous statistical and analytical results. We compute
rotation rates by assuming conservation of angular momentum around the
center of mass at impact and that merged planetesimals relax to
spherical shapes. At the end of our simulations, the majority of
bodies that have undergone at least one merger are rotating faster
than the breakup frequency. This implies that the assumption of
completely inelastic collisions (perfect accretion), which is made in
most simulations of planetary growth at sizes 1 km and above, is
inappropriate.  Our simulations reveal that, subsequent to the number of
particles in the patch having been decreased by mergers to half its initial value, the presence of larger
bodies in neighboring regions of the disk may limit the validity of
simulations employing the patch approximation.
\end{abstract}

% The different journals have different requirements for keywords.  The
% keywords.apj file, found on aas.org in the pubs/aastex-misc directory, 
% contains a list of keywords used with the ApJ and Letters.  These are 
% usually assigned by the editor, but authors may include them in their 
% manuscripts if they wish. 

\keywords{}

\section{Introduction} 

The ``planetesimal hypothesis'' states that planets grow within
circumstellar disks via pairwise accretion of small solid bodies known
as planetesimals (Chamberlin 1905; Safronov 1969; Hayashi \etal 1977).
The process of planetary growth is generally divided for convenience
and tractability into several distinct stages.  In the first stage,
microscopic grains collide and grow via pairwise collisions while
settling towards the midplane of the disk.  If the disk is laminar,
then the solids may collapse into a layer that is thin enough for
gravitational instabilities to occur (Edgeworth 1949; Safronov 1960;
Goldreich and Ward 1973; Youdin and Shu 2002; Garaud and Lin 2004;
Johansen \etal 2007); such instabilities would have produced
planetesimals of $\sim$ 1 km radius at 1 AU from the Sun.  If the disk
is turbulent, then gravitational instabilities may be suppressed because
the dusty layer remains too thick.  Under such circumstances,
continued growth via binary agglomeration depends upon (currently
unknown) sticking and disruption probabilities for collisions between
larger grains (Weidenschilling and Cuzzi 1993; Weidenschilling 1995).
The gaseous component of the protoplanetary disk plays an important
role in this stage of planetary growth (Adachi \etal 1976;
Weidenschilling 1977; Rafikov 2004).
 
Once solid bodies reach kilometer-size (in the case of the
terrestrial region of the proto-solar disk), gravitational
interactions between pairs of solid planetesimals provide the dominant
perturbation of their basic Keplerian orbits.  Electromagnetic forces,
collective gravitational effects, and in most circumstances gas drag,
play minor roles.  These planetesimals continue to agglomerate via
pairwise mergers.  The rate of solid body accretion by a planetesimal
or planetary embryo is determined by the size and mass of the
planetesimal/planetary embryo, the surface density of planetesimals,
and the distribution of planetesimal velocities relative to the
accreting body.  The evolution of the planetesimal size distribution
is determined by the gravitationally enhanced collision cross-section,
which favors collisions between bodies having smaller relative
speeds.  Runaway growth of the largest planetesimal in each
accretion zone appears to be a likely outcome.  The subsequent
accumulation of the resulting planetary embryos leads to a large
degree of radial mixing in the terrestrial planet region, with giant
impacts probable.  Growth via binary collisions proceeds until the
protoplanets become dynamically isolated from one another (Lissauer 1987, 1995).

Numerous groups have attempted to examine the accumulation and
dynamics of 1 km planetesimals via numerical simulations. The
statistical approach (Greenberg \etal 1978; Kolvoord and Greenberg
1992) came to be known as the particle-in-a-box (PIAB) method. More
recent PIAB investigations, also beginning with 1 km planetesimals,
have been performed by Weidenschilling \etal (1997) and Kenyon and
Bromeley (2006). PIAB assumes that the velocity distribution of
planetesimals is a smooth function. Each particle effectively sees a
``sea'' of particles. The advantage of this model is that one need not
sum the gravity between all the bodies; rather, the relative
velocities and impact parameters of two colliding particles can be
randomly selected from a simple function.
 
Lecar and Aarseth (1986) were the first to apply $N$-body modeling of
planetary growth with a simulation of 100 lunar-sized bodies that
interacted directly and accreted.  Computer power and algorithm
sophistication increased through the 1990's, but these improvements
were not sufficient to enable the direct $N$-body modeling of growth
from 1 km planetesimals. Nevertheless, great strides were made toward
understanding the final stages of planet formation (\eg Agnor
\etal 1999; Chambers 2001; Kokubo and Ida 2002; Raymond \etal 2006; O'Brien \etal 2006; Morishima \etal 2008). The highest $N$ (the number of particles) simulation to date is that of Richardson \etal (2000), which modeled one million 150 km radius
particles for $10^3$ years.
 
In this investigation we simulate 1 km planetesimal
growth with an $N$-body model. Our approach is not to consider an entire
disk of planetesimals, as calculating gravitational interactions
between more than 1 trillion particles is an intractable task for the
foreseeable future. Instead we focus on small, square, shearing
patches of the disk, containing up to $10^5$ particles. In this way,
we follow growth over several orders of magnitude, compare
self-consistent and statistical calculations, and lay the groundwork
for future investigations of direct simulations of planetesimal
growth.

In $\S$2, we describe the numerical integration techniques. In $\S$3,
we summarize the initial conditions of our simulations. In $\S$4, we
present the results for our baseline model, in which the magnitude of
the initial velocity dispersion is equal to the escape speed of 1 km
planetesimals. In $\S$5, we compare the results of two simulations
with different initial velocity dispersions. In $\S$6, we discuss some
of the key results to emerge from these simulations. Finally, in
$\S$7, we draw more general conclusions, extrapolate our results to
longer times and larger orbital radii, and describe future directions
of research. Appendix A lists all symbols and abbreviations used in
this article. Appendix B reviews an analytic method for approximating
planetesimal accumulation. Appendix C presents three of our simulations to 2000 orbits; these results have been relegated to an appendix because they probably suffer from systematic errors due to the small physical
size of the region being simulated combined with the flat slope of the size-frequency distribution of planetesimals at this epoch.

\section{Numerical Techniques}
We use the code PKDGRAV (Richardson \etal 2000; Stadel 2001; Wadsley
\etal 2004) to perform the integrations. This is a parallel, highly
scalable $N$-body algorithm, originally designed for cosmological
simulations (see, \eg Moore \etal 1998). The code incorporates
features such as multipole expansions, multistepping, and binary
merging to increase speed (described below). Despite these
sophisticated techniques, PKDGRAV cannot integrate more than $\sim
10^5$ colliding 1 km planetesimals for $10^3$ years in a reasonable
amount of CPU time with our model. Since the collision timescale is of
order minutes (see $\S$2.3), our ``baseline model'', presented in
$\S$4, required over $4 \times 10^8$ timesteps. We describe the
equations of motion of the patch in $\S$2.1, our collisional model in
$\S$2.2, the basic principles of PKDGRAV in $\S$2.3, the numerical
approximations necessary to complete these simulations in $\S$2.4, and
our methods of verification in $\S$2.5.

\subsection{Equations of Motion}
Although carrying out the simulations in a patch greatly reduces $N$,
it adds the new complication of simulating Keplerian shear: we must
account for the gradient in the Sun's gravitational field and the geometry of the disk. If we
examine small patches of the disk such that $W \ll r$, where $W$ is the
width and length of the square patches and $r$ is the distance from the Sun, then we may approximate
quantities, such as surface density, scale height, and circular
velocity, as varying linearly in the patch.

Two coordinate systems are used in this model: heliocentric cylindrical and
locally Cartesian. The heliocentric system is the more physically meaningful
system, as it is based on the geometry of the disk. In these
coordinates, the origin is located at the position of the
central star, and ($r$, $\theta$, $z$) have their standard meanings. We
implement the Cartesian system inside the patch, with the origin
at the center of the patch. For small $\theta$, these two coordinate systems are related
by the following expressions:
\begin{equation}
\label{eq:coords}
\begin{array}{l}
r = r_{patch} + x,\\
\theta = \frac{3}{2}\Omega_{patch}\frac{x}{r}t + \frac{y}{r},\\
\end{array}
\end{equation}
where $r_{patch}$ is the heliocentric radius of the center of the
patch and $\Omega_{patch}$ is the Keplerian orbital frequency of the patch. Essentially the radial
direction is mimicked by $x$ and the tangential direction by
$y$. Similarly, the velocities are related by
\begin{equation}
\label{eq:velocities}
\begin{array}{l}
\dot{R} = \dot{x},\\
\dot{\theta} = \frac{3}{2}\Omega_{patch}\frac{x}{R} + \frac{\dot{y}}{R}.\\
\end{array}
\end{equation}
The $z$-velocities and positions are identical in the two
coordinate systems.

For the purpose of the integration, we use the Cartesian system,
as outlined in Wisdom and Tremaine (1988), which was based on Hill (1878). The equations of motion
inside the patch are
\begin{equation}
\label{eq:motion}
\begin{array}{l}
\ddot{x} - 2\Omega_{patch}\dot{y} - 3\Omega_{patch}^2 x = -\frac{\partial\phi}{\partial x}\\
\ddot{y} + 2\Omega_{patch}\dot{x} = -\frac{\partial\phi}{\partial y}\\
\ddot{z} + \Omega_{patch}^2z = -\frac{\partial\phi}{\partial z},
\end{array}
\end{equation}
where $\phi$ represents the interparticle potentials in the disk. Wisdom and Tremaine assume massless particles ($\nabla\phi = 0$), and provide the general solution
\begin{equation}
\label{eq:soln}
\begin{array}{l}
x = x_g + A\cos (\Omega_{patch}t) + B\sin (\Omega_{patch}t),\\
y = y_g - \frac{3}{2}\Omega_{patch}x_gt - 2A\sin (\Omega_{patch}t) + 2B\cos (\Omega_{patch}t),\\
z = C\cos (\Omega_{patch}t) +D\sin (\Omega_{patch}t),
\end{array}
\end{equation}
where $A, B, C$, and $D$ are arbitrary constants that correspond to
the amplitude of the excursions from a circular, planar orbit, and
$x_g$ and $y_g$ are the guiding centers of the epicycles of the
particles. The $\frac{3}{2}\Omega_{patch}x_g$ term is known as the
``shear'' and represents the Keplerian motion relative to that at
$x=0$. For negative $x$, the shear is positive, as these particles are
closer to the star. The $y$-motion is the tangential motion relative to the center of the patch, not the
azimuthal motion of the patch as a whole.

The equations of motion are integrated with a second-order generalized
leapfrog scheme derived via the operator splitting formalism of Saha
and Tremaine (1992). For the case of orbits in
the patch, this integrator prevents secular changes in the guiding
center of the epicycle.

\subsection{Collision Model}
Our model assumes perfect accretion. In our implementation,
collisions result in one perfectly spherical particle, and all particles have the same
density. To conserve angular
momentum, the resultant particle receives the net angular momentum
of the two colliders (spin plus motion relative to the center-of-mass of the colliding particles), without dissipation (although energy is decreased by 20\% during a collision). Our assumption of perfect accretion allows particles to spin faster than break-up, the latter being given by
\begin{equation}
\label{eq:Pmin}
p_{min} = \sqrt{3\pi/G\rho_{pl}}.
\end{equation}
(For a density of 3 g/cm$^3$, $p_{min} = 1.9$ h.) Real gravitational aggregates do not merge completely under such circumstances, and hence this assumption will tend to artificially increase planetesimal masses. Conversely, spheres have the smallest
possible cross-section for a given volume, so our algorithm is
suppressing growth with this approximation. Therefore these two
features of our model counteract each other, although we cannot say to
what degree they cancel each other out.

The assumption that the particles are always gravitational aggregates may 
break down during high-energy collisions, in which case melting may occur. 
For tractability, we ignore this possibility. We also ignore any potential 
radiogenic heating, \eg by $^{26}$Al, as the melting timescale is much 
longer than the collisional timescale.

\subsection{PKDGRAV}
PKDGRAV is very efficient because it calculates gravitational forces
rapidly.  The code works by recursively dividing the physical space, the
``domain'', into smaller cells that are organized in a tree.  The
gravitational forces are then calculated by traversing this tree
(Barnes and Hut 1986).  To calculate the forces on a particle from a given
cell, the code determines the apparent size of the cell, that is the
angle the cell subtends as viewed from the particle.  If this angle is
smaller than some minimum angle, $\Theta$, then all moments up to
hexadecapole of the particles in the cell are used to calculate the force.
Otherwise the cell is opened and the test is repeated on the
subcells.  This approximation changes the speed of the $N$-body
calculation from $O(N^2)$ to $O(N\log N)$.  For further details see
Stadel (2001) and Wadsley \etal (2004).  For the simulations
described here, we used $\Theta = 0.7$ radians.

PKDGRAV also permits the use of periodic boundary conditions, which are
required to prevent the patch from self-collapsing. To do this, each
patch is reproduced 8 times, in the form of ``ghost cells'' that
completely surround the actual patch in the $x$ and $y$
directions. Eq.\ (\ref{eq:soln}), however, also requires that
particles in the ghost cells move at the appropriate shear
velocity. The centers of the ghost cells therefore move in the
$y$-direction as prescribed by Eq.\ (\ref{eq:soln}).

Gravity calculations per time interval are minimized by a
multistepping algorithm. Multistepping divides the base timestep into
smaller intervals. For our simulations the timestep is based on the
largest acceleration a particle feels at each timestep. In this
algorithm isolated planetesimals move at the base timestep,
$t_{base}$, set to about 100 steps per orbit. As particles approach
each other their timesteps drop to a minimum value, until they either miss
or collide. Each smaller timestep is a factor of 2 shorter then the
previous in order to keep all of the base gravitational kicks
commensurate. Typically 95\% of particles are on the longest timestep,
but we still resolve all collisions.

Two timescales are relevant in this problem, the orbital time and the
crossing time of two particles that just miss. The orbital
period, $P$, is of order one year, but the crossing time is of
order minutes. For two particles of radius $R$ to just miss, on a parabolic orbit,
the crossing time is
\begin{equation}
\label{eq:tcross}
t_{cross} = r/v_{esc} = \frac{R}{\sqrt{\frac{8\pi}{3}G\rho R^2}} = \sqrt{\frac{3}{8\pi G\rho}},
\end{equation}
and is therefore independent of the masses. In this scenario and with densities of 3 g/cm$^3$, the
crossing time is 772 s. $P$ and $t_{cross}$ determine $t_{base}$ and
$t_{min}$, the maximum and minimum possible timesteps. To be
conservative (note that the version of PKDGRAV employed in this investigation does not integrate Eq. [\ref{eq:motion}] symplectically), we set these values as
\begin{equation}
\label{eq:tbase}
t_{base} = \eta P\\
\end{equation}
and
\begin{equation}
\label{eq:tmin}
t_{min} = \eta t_{cross},
\end{equation}
where $\eta$ is a scale factor, chosen such that the integrals of motion are constant to a satisfactory degree. Convergence tests showed that for the systems we consider, $\eta = 1/300$ provided the necessary accuracy. The number of available timesteps, $\zeta$, is
\begin{equation}
\label{eq:rung}
\zeta = 1 + \textrm{log}_2\Big(\frac{t_{base}}{t_{min}}\Big).
\end{equation}
At 0.4 AU  and $\rho = 3$ g/cm$^3$, this translates to $\zeta = 14$.

\subsection{Numerical Approximations}
Although PKDGRAV is fast relative to other integration techniques, we
must make several additional approximations in order for the
simulation to be practical, and to appropriately model Eq.\
(\ref{eq:soln}). These approximations do not alter the physics
appreciably, and resulted in nearly a factor of 10 speed-up. 
 
As mentioned above, the timestep for each particle is determined by its current acceleration, which is a function of the local mass density. Particles on large timesteps are not close to
other particles and cannot be close to collision. We
therefore set a ceiling for timesteps to search for
collisions. This small
modification increases speed by a factor of 2 -- 3 depending on the surface density of
the patch (and hence the local density).

Occasionally during the evolution of the patches, binaries
(gravitationally bound pairs of particles) form. This is a natural
result of 3-body encounters. Although it is preferable to allow these
binary systems to evolve normally, they often become stuck in the
shortest timesteps, greatly diminishing the advantage of 
multistepping. Therefore, we artificially merge them. In our simulations, binary merging accounts for about 0.01\% of all merging events and therefore this small change in the total energy of the
simulation should be negligible. Simulations of globular clusters have
shown that tight binaries tend to get tighter during a 3-body
encounter, and loose binaries tend to become looser (Heggie 1975). To
make binary merging as realistic as possible we therefore also require
the eccentricity of the particles involved to be less than 0.9, so that
wide binaries may still be disrupted.

\subsection{Verification}
Given the complexity of this problem (non-inertial frame, large range
of growth, and numerical approximations), we need to quantify
the accuracy of our methodology. The patch framework provides some inherent tests, but we must
also understand the statistics of the problem; this patch is supposed to be a representative piece of a much larger
annulus of material. At some point, the number of particles drops to a
sufficiently small number that the results cannot be trusted. In this
subsection we describe our methodology for verifying the results presented in
$\S\S$4 -- 5.

In the shearing model, the particles are in a non-inertial
frame. Therefore the integrals of motion normally associated with
dynamics (momentum, energy, angular momentum), are not directly
applicable. In this formalism there are two constants of motion we
use to check the validity of these simulations. In the
nomenclature of Wisdom and Tremaine (1988), these conserved integrals
are
\begin{equation}
\label{eq:uw}
\begin{array}{l}
u \equiv \frac{\sum_{l=1}^{N}{m_l}\frac{dx_l}{dt}}{M_{patch}}\\
w \equiv \frac{\sum_{l=1}^{N}{m_l}(\frac{dy_l}{dt}+\frac{3}{2}\Omega_{patch}x_l)}{M_{patch}},
\end{array}
\end{equation}
where $M_{patch}$ is the total mass in the patch. These parameters
essentially correspond to the center of mass velocity. For this situation,
the center of mass should remain motionless; we need to verify that
$u$ and $w$ remain much less than the random velocities. Note that our
definition includes the masses of the particles, whereas the Wisdom and
Tremaine (1988) definition did not, as they weighted all particles equally.

Eq.\ (\ref{eq:uw}) would be sufficient if we were integrating
particles without growth, which we are not. The patch model at least requires that no particle is
dominant. Therefore a zeroeth order requirement is that no particle
reaches a mass equal to that of the sum of the remaining particles. This
constraint, however, fails to take into account how the most massive
particles alter the dynamics of the disk. As a typical particle
passes by the largest mass particle, it receives a kick, and energy
associated with Keplerian motion may be transformed into the random
motions of the swarm, increasing the velocity dispersion, \ie
``viscous stirring'' (Wetherill and Stewart 1989). Afterward that
increase in random velocity should be damped down by subsequent
interactions with other typical particles. To verify that no particles grow so large as to dominate the stirring in the patch, we consider the ratio of the gravitational stirring of the largest particle to the sum of all other particles:
\begin{equation}
\label{eq:stir}
S \equiv \frac{m_{max}^2}{\sum_{l\le l_{max}}N_lm_l^2 - m_{max}^2}
\end{equation}
(Lissauer \& Stewart 1993). Exactly how large $S$ can grow is not clear \textit{a priori}, but values below $\sim 0.1$ should be satisfactory.

Should the mass distribution in the patch suggest that
significantly massive planetesimals are present in the disk, but not
present in a patch, then the patch is not modeling a large enough
region of the disk. We check this possibility by plotting $m^2N_k$ as
a function of $m$, where $N_k$ is the number of particles in mass bin
$k \equiv m/m_1$. We will see in $\S$4.4.2 and Appendix C that unmodeled large bodies are likely to become a problem as our model evolves.

A final point of concern is the size of the epicycles of particles
compared to the size of the patch. Should the eccentricity of a
particle grow large enough that the radial excursions ($2ae$, where $a$ is
semi-major axis, $e$ is eccentricity) exceed the size of the patch, then we may not be sampling the region appropriately.  We parameterize this effect as
\begin{equation}
\label{eq:beta}
\beta \equiv \frac{2r_{patch}e}{W},
\end{equation}
where $W$ is the patch width. The equations of motion do not depend on
eccentricity, so there are no numerical problems if this occurs,
only concerns about the physical interpretation of this
phenomenon. Our choice of $W$ is very
small compared to the size of the terrestrial annulus. As long as
$\beta$ remains less than or close to unity, the radial mixing
throughout the disk is negligible, and our interpretations are
independent of the particles' eccentricity. Only if $\beta$ grows to
large values is there cause for alarm. We focus on the
second largest $\beta$ value, as occasional strong kicks could
temporarily result in large eccentricities for one particle. 

\section{Initial Conditions}
We assume a surface density of 37.2 g/cm$^2$ at 0.4 AU, 30\% higher than predicted by the minimum mass solar
nebula model (Hayashi 1981), but consistent with previous studies (Richardson \etal 2000). The bulk density of the planetesimals, $\rho_{pl}$, is taken to be 3
g/cm$^3$. The mass of a planetesimal is thus
\begin{equation}
\label{eq:m_pl}
m_{pl} = \frac{4\pi}{3}\rho_{pl}R^3 = 1.26 \times 10^{16} \textrm{g}.
\end{equation}
Assuming the surface density of the disk scales as $\Sigma \propto a^{-1}$, the number of planetesimals in the annulus 0.4 AU $\le a \le$ 2.5 AU 
is 
\begin{equation}
\label{eq:N}
N = \frac{M_{ann}}{m_{pl}} = 3.5 \times 10^{12},
\end{equation}
As we are unable to directly integrate
trillions of particles, we must reduce $N$ to a tractable
value by dividing the disk into patches.

Most of our simulations begin with the root mean squared (RMS) velocity dispersion, $v_{RMS}$, set to equal to the escape speed of 1 km
planetesimals, $v_{esc}$ (Safronov 1969; Stewart and Wetherill
1988). This relationship assumes that the system is relaxed, which may
or may not be the case depending on the processes and timescale to
form planetesimals. This RMS speed is the magnitude of the random
motion of the particles, but it is not distributed isotropically due
to the 2:1 axis ratio of the epicycle; instead the distributions are
described by
\begin{equation}
\begin{array}{l}
v_x = \sqrt{\frac{2}{3}}v_{esc}\\
v_y = v_z = \frac{1}{\sqrt{6}}v_{esc}
\end{array}
\end{equation}
(Binney and Tremaine 1994). For a radius of 1 km and $\rho_{pl} = 3$
g/cm$^3$ particles, $v_{esc}$ is 1.29 m/s.

From this velocity distribution, we calculate the equilibrium
vertical density profile of the disk. To do this we assume the
disk is in a state of hydrostatic equilibrium; the ``pressure'' of the
vertical component of the velocity distribution maintains its
thickness. The vertical component of solar gravity increases linearly
with distance from the midplane, therefore the density follows a
Gaussian profile,
\begin{equation}
\label{eq:rho}
\rho = \rho_0e^{-\frac{z^2}{2Z_0^2}},
\end{equation}
where $\rho$ is the density, $\rho_0$ is the density at the midplane,
and $Z_0$ is the scale height of the disk. The (Gaussian) scale height
is determined by the $z$-velocity distribution, the mass of the
central star and the orbital radius. Hydrostatic equilibrium allows a
calculation of the scale height:
\begin{equation}
\label{eq:Z0}
Z_0 = \sqrt{\frac{v_z^2r^3}{2GM_\odot}} = \sqrt{\frac{v_z^2}{2\Omega_z^2}}.
\end{equation}
The scale height also depends on the self-gravity of the planetesimal
disk. To compensate for this, we implement a ``vertical frequency
enhancement'' to simulate an infinite plane sheet of matter so that 
\begin{equation}
\label{eq:rhoz}
\Omega_z = \Omega_{patch} + \sqrt{\frac{2\pi GM_{patch}}{W^3}}.
\end{equation}
The second term in Eq.\ (\ref{eq:rhoz}) is an analytic restoring force that simulates the disk's
self-gravity and reduces $Z_0$. For our baseline model (Simulation L$_1$ presented in $\S$4), $\Omega_z$ is 0.2\% 
larger than $\Omega_{patch}$. 

Our nomenclature for simulations is based on the relative size
and the initial dynamical state. L stands for ``large'', M for
``medium'' and S for ``small''. The subscript is the ratio of the
initial velocity dispersion to the escape speed of a 1 km planetesimal. In Table 1, we present the
initial conditions for these simulations.

We assume that scattering is very effective, and that gas drag is negligible. Simulation L$_1$, M$_1$ and S$_1$ presume the planetesimals are in a form of
equilibrium. However, there remain many unknowns in the formation of
these planetesimals and variations of a factor of 2 are plausible. We
therefore ran two integrations with non-equilibrium
initial conditions. We integrated one system that began with $v_{RMS}
= 0.5v_{esc}$ (Simulation M$_{0.5}$), and one with $v_{RMS} =
2v_{esc}$ (Simulation M$_2$). In these simulations the scale height
was set by Eq.\ (\ref{eq:Z0}). These two simulations are presented in
$\S$5.

Each of the simulations has been run at least until the
number of bodies in the patch has been halved, $t_{1/2}$. This time is
related to the mean free time, $\tau$, between collisions, which can
be approximated as
\begin{equation}
\label{eq:tau}
\tau \approx \frac{m_1}{\rho_0\sigma_{pl}v_{RMS}} \approx t_{1/2},
\end{equation}
where $\sigma_{pl}$ is the gravitational cross-section, $\sigma(1+v^2_{esc}/v^2_{RMS})$, and $\rho_0$
is given by
\begin{equation}
\label{eq:rho0}
\rho_0 = \frac{\Sigma}{\sqrt{2\pi}Z_0}.
\end{equation}
Eq.\ (\ref{eq:tau}) is only valid at the midplane at time $t = 0$. The
density changes with $z$, and $v_{RMS}$ change with time. Nonetheless,
this simple model provides valuable insights into the dynamics. For the initial conditions we use, $\tau \approx  500$ orbits.

Richardson \etal (2000) modeled a planetesimal disk with $10^6$
particles, but they simulated a later stage of planetary growth in
which the collision rate was much lower than it is here. Therefore,
our integrations proceed slower, even with the advanced
methodologies described $\S$2.3. If we set $N_{patch} \sim 10^5$, then
the simulations are large, but we may still examine several different
initial conditions. We choose a square patch of width $W =
10^{-3}r_{patch}$ for our largest simulation, L$_1$. In our model,
this choice corresponds to an initial $N$ of 106,130, a small enough
number to be tractable, but large enough to be statistically
significant.  Note, however, that the medium-sized simulations have
dimensions $W = 5\times 10^{-4}r$, and the small one $W = 2.5\times
10^{-4}r$. These cases run faster, and are included to test various
assumptions in the baseline case.

\section{The Baseline Model}
The simulations of our baseline model (those with subscript 1) begin
with 1 km radius particles with an RMS velocity equal to that of the
escape speed of 1 km particles with a density of 3 g/cm$^3$. From this
dispersion the initial scale height is set by Eq. (\ref{eq:Z0}). The
simulation with the most particles, L$_1$, required 354 orbits (89.6
years) to halve the total number of particles. The
M$_1$ and S$_1$ simulations required 349 and 361 orbits, respectively,
to reach $t_{1/2}$. In this section we describe the results of these
three simulations. Table 2 summarizes some of the results of our
simulations at $t_{1/2}$. At this time,
$t_{1/2}$, the patch is nearing a condition in which it is too small
to correctly model the velocity distribution (see $\S$4.4.2); results of these simulations beyond $t_{1/2}$ are presented in Appendix C.

\subsection{Mass Growth}
The mass spectra of Simulations
L$_1$, M$_1$, and S$_1$ at 100, 200, 300 and 354 orbits ($t_{1/2}$ for
L$_1$) are plotted in Fig.\ \ref{fig:eq-masslog}.  Figure \ref{fig:eq-mloglog} shows the (differential) mass distributions with
logarithmic bins of size $2^lm_1, l=1,2,3,...$, \ie the first bin
contains all particles of mass $m_1$, the second bin contains
particles of $2m_1$ and $3m_1$, the third bin contains particles in
the range $4m_1 \le m \le 7m_1$, etc. The histograms shown in Fig.\
\ref{fig:eq-mloglog} appear to follow a power law size distribution. The one apparent
exception is the single particle in the largest bin for L$_1$ at
$t_{1/2}$, which is detached by one bin from the remaining particles
(the ``swarm''). However, upon closer inspection, this particle has a
mass of 275 $m_1$, whereas the second and third largest particles have
masses of 122 and 116 $m_1$, respectively. All three of these particles just
miss being members of the 128 -- 255 $m_1$ bin. Therefore, the distributions presented in Fig.\ \ref{fig:eq-mloglog} suggests that runaway growth has \textit{not} occurred.

\medskip
\epsfxsize=8truecm
\epsfbox{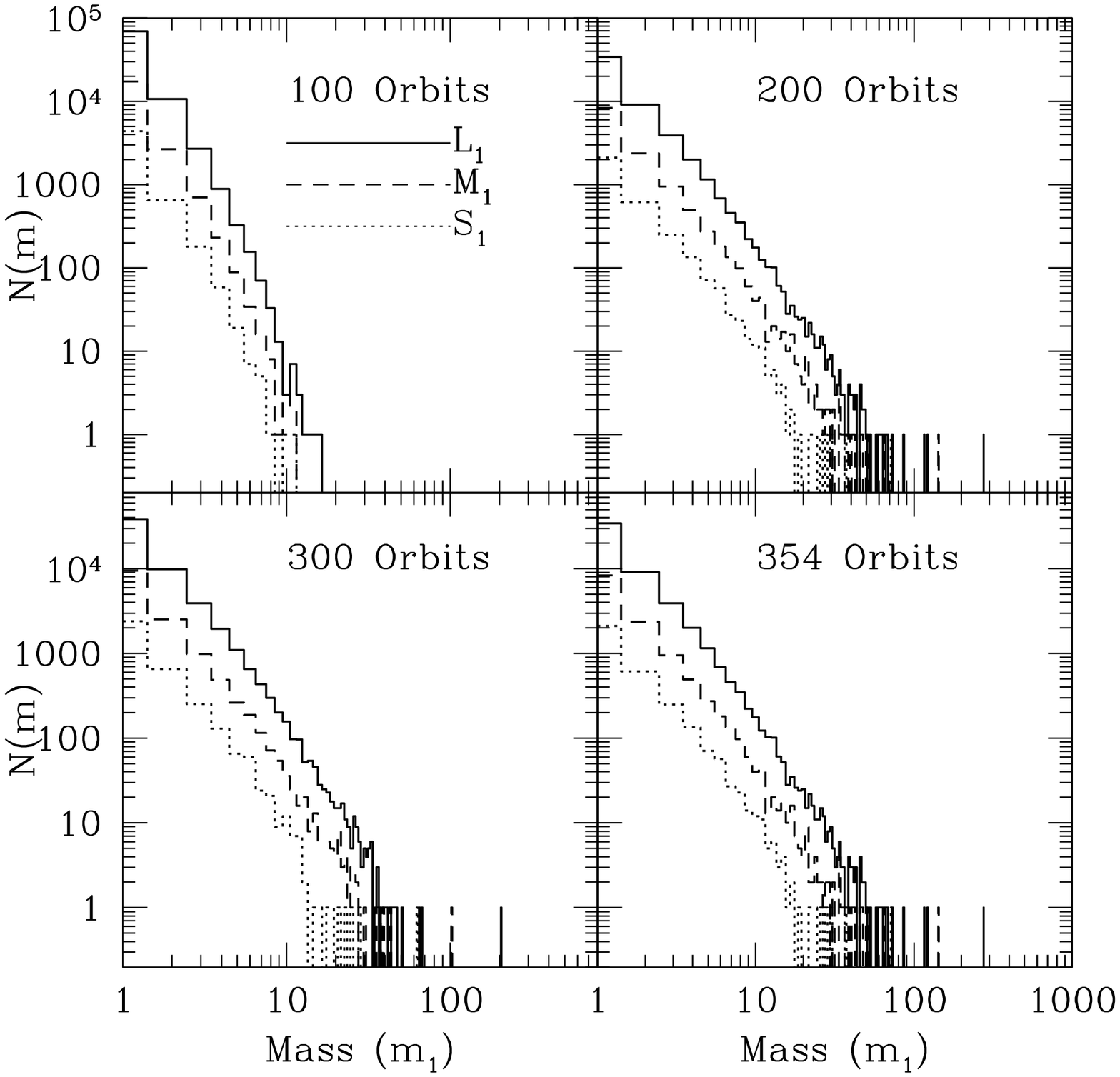}
\figcaption[]{\label{fig:eq-masslog}\small{The semi-log (differential) mass distributions for Simulations L$_1$ (solid line), M$_1$ (dashed line) and S$_1$ (dotted line) after 100 orbits (top left), 200 orbits (top right), 300 orbits (bottom left) and 354 orbits (bottom right).}}
\medskip

\medskip
\epsfxsize=8truecm
\epsfbox{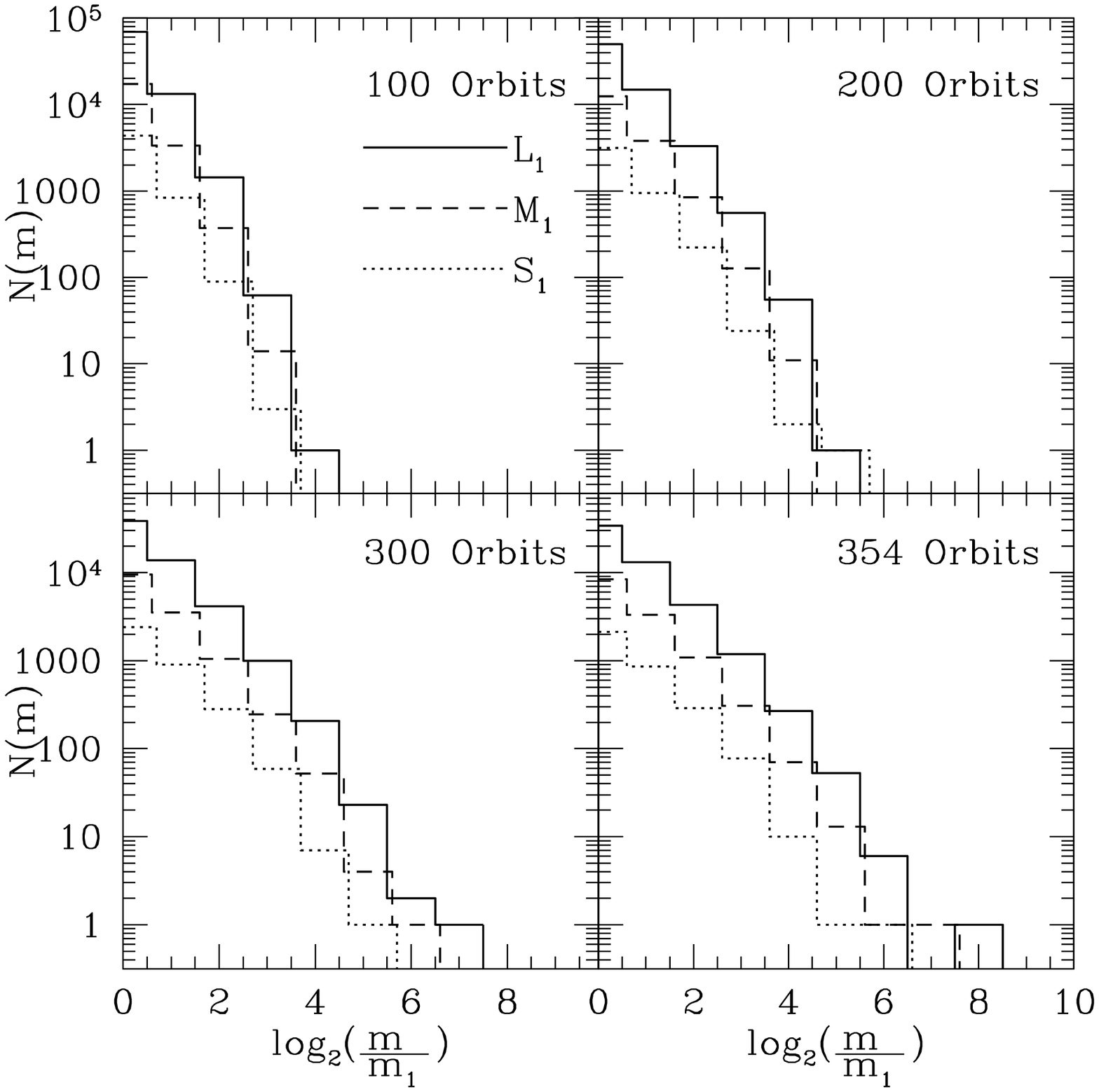}
\figcaption[]{\label{fig:eq-mloglog}\small{The log-log (differential) mass distributions of Simulations 
L$_1$ (solid line), M$_1$ (dashed line) and S$_1$ (dotted line) at 100 orbits (top left), 200 orbits (top right), 300 orbits (bottom left), and 354 orbits (bottom right). Note that the M$_1$ simulation data have been offset to the right by 0.1 and the S$_1$ data by 0.2 in order to improve readability.}} \medskip

To further examine the nature of the mass distribution, we
matched the mass distribution to single parameter fits. We consider a power law,
\begin{equation}
\label{eq:plaw}
N(k) \propto k^{-b},
\end{equation}
and an exponential, 
\begin{equation}
\label{eq:exp}
N(k) \propto e^{-k/c}.
\end{equation}
We fit the constants $b$ and $c$ by using the line method to minimize
$\chi^2$ (Press \etal 1996) with one sigma uncertainties assumed to be
$\sqrt{N_k + .75} + 1$.  For small $N_k$ this gives a better
approximation to the distribution of $\chi^2$ for sparsely sampled
data (Gehrels 1986), \eg some bins at large $k$ have zero particles.
The fits are constrained so that the total mass is the same as in the
simulation, and the number of degrees of freedom is approximately
equal to $m_{max}/m_1$.  We must take into consideration the fact
that the first bin is unusual in that all the particles were
originally in that bin. Therefore we also consider fits that exclude
the data at $m = m_1$. These fit parameters are denoted with a
prime. We found that exponential fits are always poor, so we will
focus on the power law fits.

In Fig.\ \ref{fig:eq-c2f3} we show the mass distributions of
Simulations L$_1$ (top), M$_1$ (middle) and S$_1$ (bottom), with the
analytic fits for comparison at each simulation's $t_{1/2}$. The
values of all the parameters are presented in Table 3. In that table
we see that the values of unreduced $\chi^2$ are all significantly
better when the first data point is excluded. All
fits include the large-mass particles and we conclude that these
largest particles do not represent a new class (\ie
runaway growth has not occurred). Note that forcing the fitted
distribution to contain the same total mass as the simulation forces
the exponential cases to converge to poor fits.

\medskip
\epsfxsize=8truecm
\epsfbox{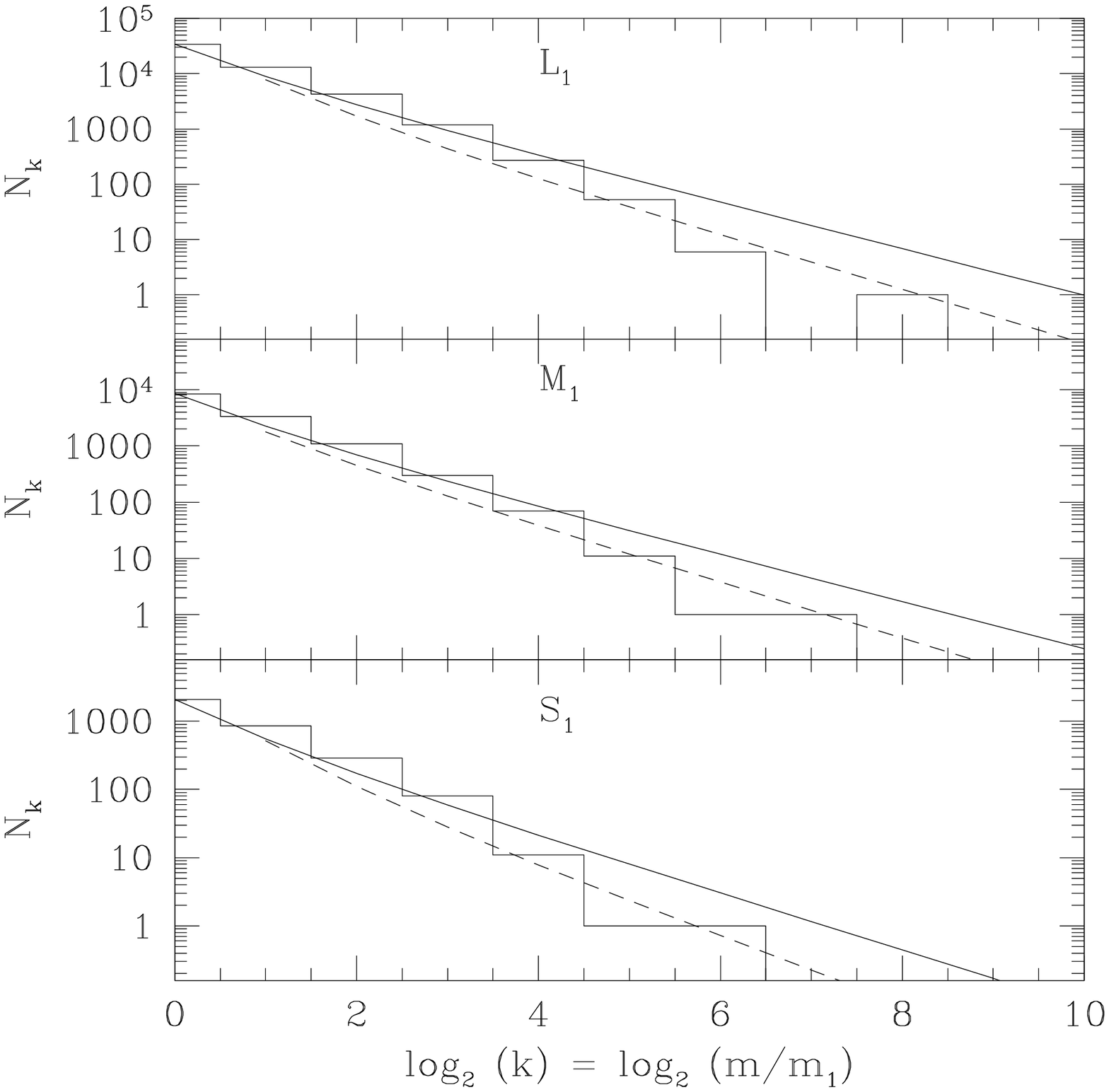}
\figcaption[]{\label{fig:eq-c2f3} \small{Comparison of the mass distributions at $t_{1/2}$ for the baseline simulations. The histograms represent the mass distribution at each model's value of $t_{1/2}$ and the straight lines are the power law fits, solid includes the first bin, dashed does not.}} 
\medskip

In Fig.\ \ref{fig:eq-merge} we compare the merger rate ($-dN/dt$) of  
Simulation L$_1$ with that predicted by the mean free time of the patch. 
The prediction comes from the following relation
\begin{equation}
\label{eq:dNdt}
\frac{dN}{dt} = -\frac{N\rho_0<\sigma_{pl}>v_{RMS}}{<m>},
\end{equation}
where the brackets indicate quantities averaged over the entire swarm. We assume the cross-section to be the gravitationally enhanced area of two particles with the average radius. The right hand side of Eq.\ (\ref{eq:dNdt}) is the
number of particles divided by the mean free time.  Initially the mean
free time argument underpredicts the merger rate and later it is about
right. Overall this simplified estimate of the merger rate remains
near the actual value. The rate is dropping because the total number
of particles in the patch is decreasing due to merging.

\medskip
\epsfxsize=8truecm
\epsfbox{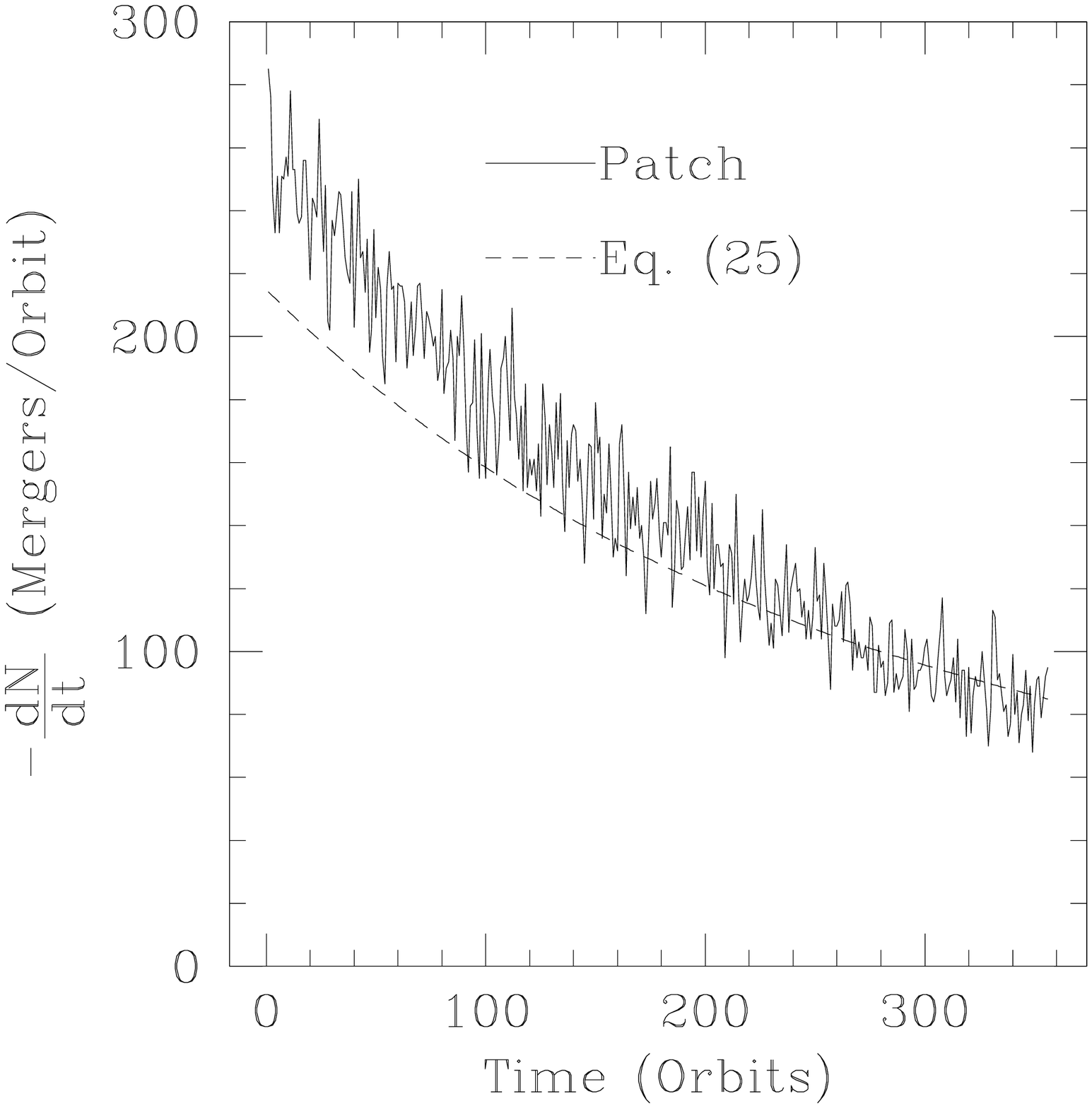}
\figcaption[]{\label{fig:eq-merge} \small{The evolution of the merger rate of Simulation L$_1$. The solid line is that observed in the patch (sampled once per orbit), and the dashed line is that predicted by the mean free time of the patch (recalculated each orbit from the instantaneous values of $v_{RMS}$, $<\sigma_{pl}>$, $\rho_0$ and $<m>$) in Simulation L$_1$.}}
\medskip

Next we compare our results to that of the constant, linear, and
product solutions to the coagulation equation (Wetherill 1990; see also App.\ B) in
Simulation L$_1$. These three solutions are relatively simple compared
to more recent derivations (see \eg Kenyon and Luu 1998; Kenyon and
Bromley 2004). However, the recent models do not provide analytic solutions
for the number of particles in each mass bin (see Eqs.\
[\ref{eq:constcoag}], [\ref{eq:lincoag}] and [\ref{eq:prodcoag}]). We find the
three collisional probability coefficients (see Eqs.\
[\ref{eq:nu1} -- \ref{eq:nu3}]) have values of $\nu_1 = 5.3
\times 10^{-8}$, $\nu_2 = 1.84 \times 10^{-8}$, and $\nu_3 = 2.65
\times 10^{-8}$ for Simulation L$_1$.

In Fig.\ \ref{fig:eq-lm} we plot the growth of some of the largest
masses in Simulation L$_1$ through 354 orbits. Also shown are the
predicted largest masses from the three solutions to the coagulation
equation, with collisional probabilities that are constant as a
function of time, and equal to the values listed above. The growth in
our $N$-body model follows the product solution to the coagulation
equation for about 250 orbits, but then the two diverge as the
$N$-body model predicts faster growth. This divergence is probably a
result of the product solution's failure to conserve mass for $t >
t_{1/2}$. At $t_{1/2}$ the largest particle has a mass of 276
$m_1$. Nonetheless, the agreement over the majority of the simulation
suggests that even the Wetherill (1990) product coagulation model is a
reasonable representation of early growth of planetesimals.

\medskip
\epsfxsize=8truecm
\epsfbox{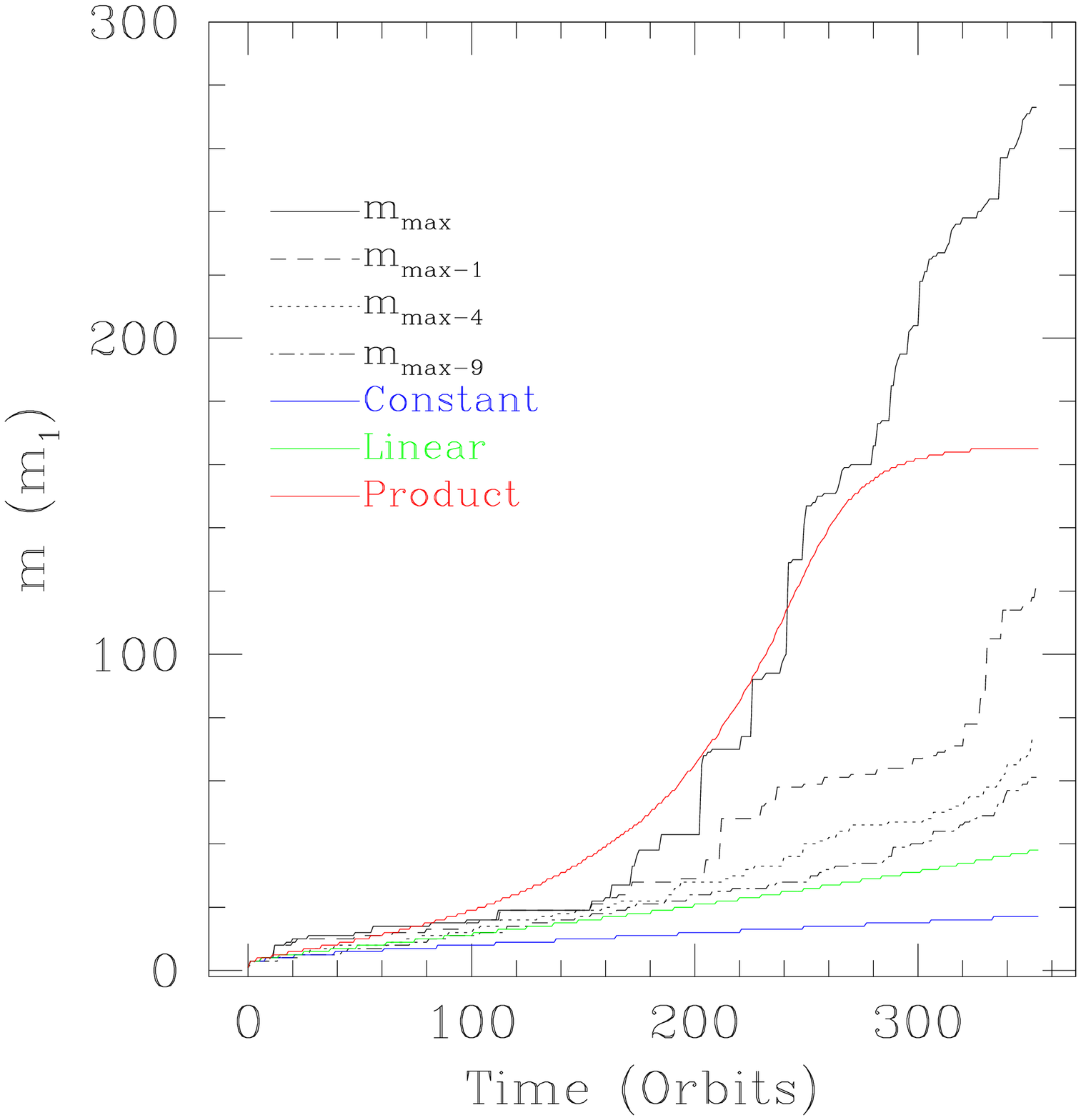}
\figcaption[]{\label{fig:eq-lm}\small{The evolution of the largest, second largest, fifth largest and tenth largest masses in Simulation L$_1$. The solid red line corresponds to $m_{max}$ in the product solution to the coagulation equation, green to the linear, and blue to the constant.}}
\medskip

In Fig.\ \ref{fig:eq-krange} we further compare Simulation L$_1$ to
the Wetherill (1990) product model. The six panels examine six different
ranges of $k$. For mass ranges, the predicted number of particles is
determined by summing all mass bins in that range. From this figure we
see that no solution to the coagulation fits the data over all values
of $k$. At low masses ($k
\lsim$ 10), the linear solution is the best-fit to the data, but at
larger masses the product solution is better. However, at these larger values of $k$, the product solution still differs from the actual distribution by more than a factor of 2.

\medskip
\epsfxsize=8truecm
\epsfbox{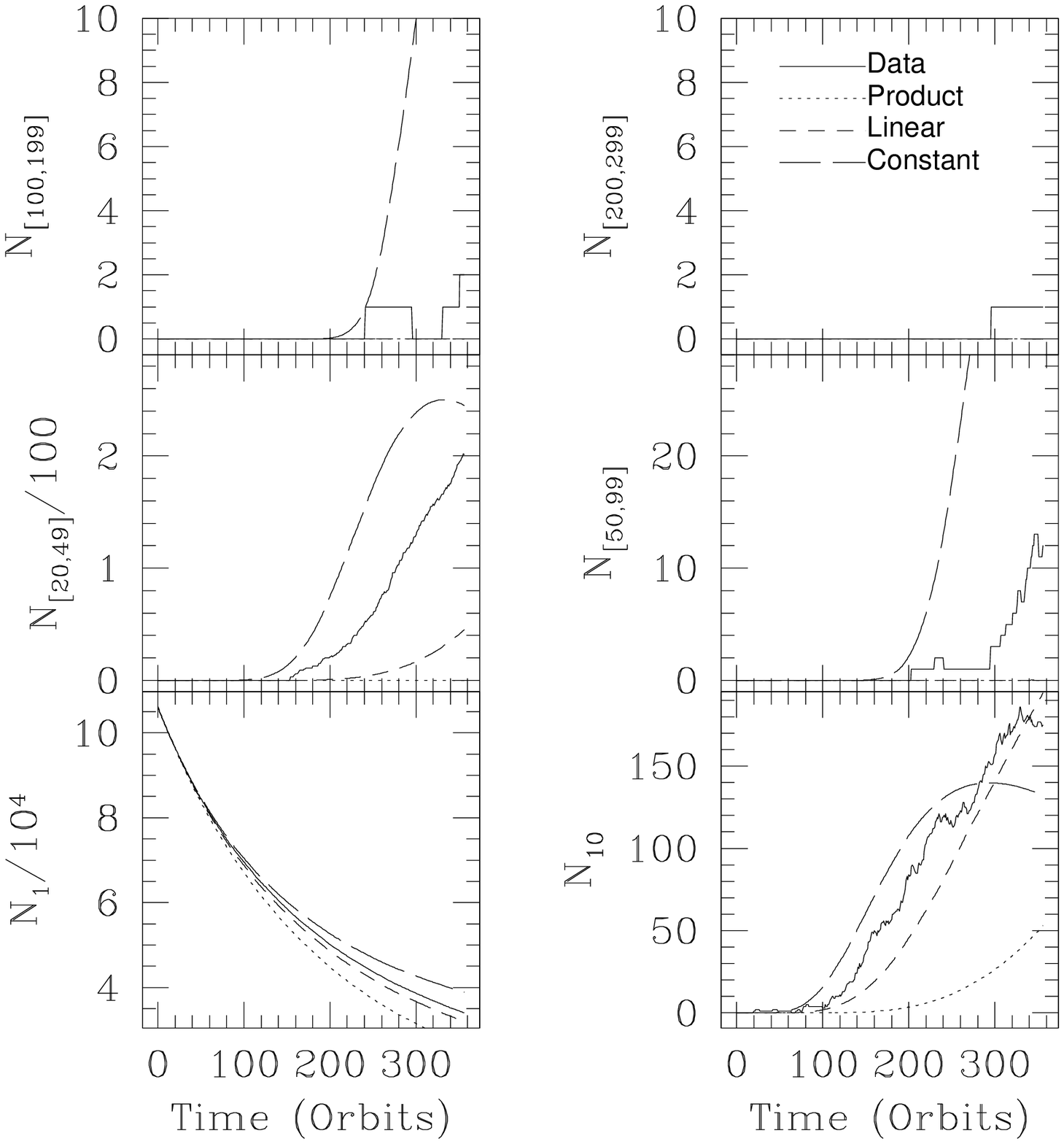}
\figcaption[]{\label{fig:eq-krange} \small{A comparison of the 
distribution of masses in Simulation L$_1$ with those predicted by 
coagulation theory for 6 different mass ranges: $k=1$ (bottom left), 
$k=10$ (middle left) $20 \le k < 50$ (middle left), $50 \le k < 100$ 
(middle right), $100 \le k < 200$ (top left) and $200 \le k < 300$ (top 
right).}}
\medskip

\subsection{Velocity Dispersion}
Figure \ref{fig:eq-vrms} shows the evolution of the RMS velocity of all
particles, the escape speed of the largest particle, and the escape
speed of the average-mass particle in Simulation L$_1$. The velocity
dispersion slowly grows to a value of $\sim$ 2 m s$^{-1}$ at $t_{1/2}$ (which is evidence of viscous stirring), while
the escape speed of the largest particle grows to a value of 8.4
m s$^{-1}$. At $t_{1/2}$ the largest body has a mass of $m_{max} = 276 m_1$
and a radius of 6.5 km.  From these values we can identify the
gravitational focusing factor for the largest body, the ratio of its gravitational cross-section to its
geometrical cross-section. This
factor is
\begin{equation}
\label{eq:gravfocus}
F_g = 1 + \frac{2Gm_{max}}{v_{RMS}^2R_{max}} = 1 + \big(\frac{v_{esc}}{v_{RMS}}\big)^2. 
\end{equation}
The final values of $F_g$ for Simulations L$_1$, M$_1$, and S$_1$ are 20.2, 13.7, and 9.2, respectively.
\medskip
\epsfxsize=8truecm
\epsfbox{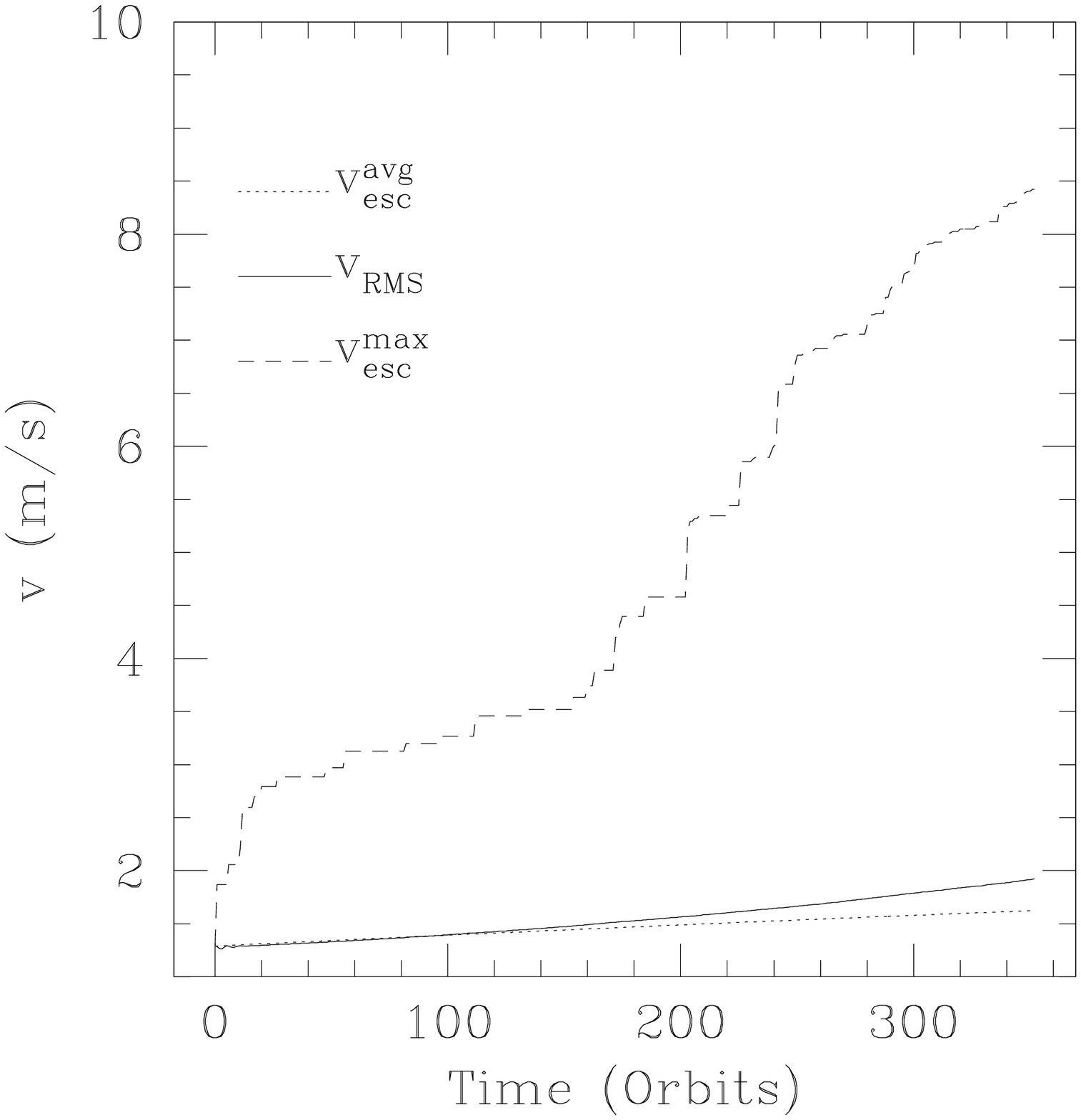}
\figcaption{\label{fig:eq-vrms} \small{The evolution of the RMS velocity of the patch (solid line),
the escape speed of the largest particle (dashed
line) and the escape speed of the mean particle (dotted line) in Simulation L$_1$. }}
\medskip

\medskip
\epsfxsize=8truecm
\epsfbox{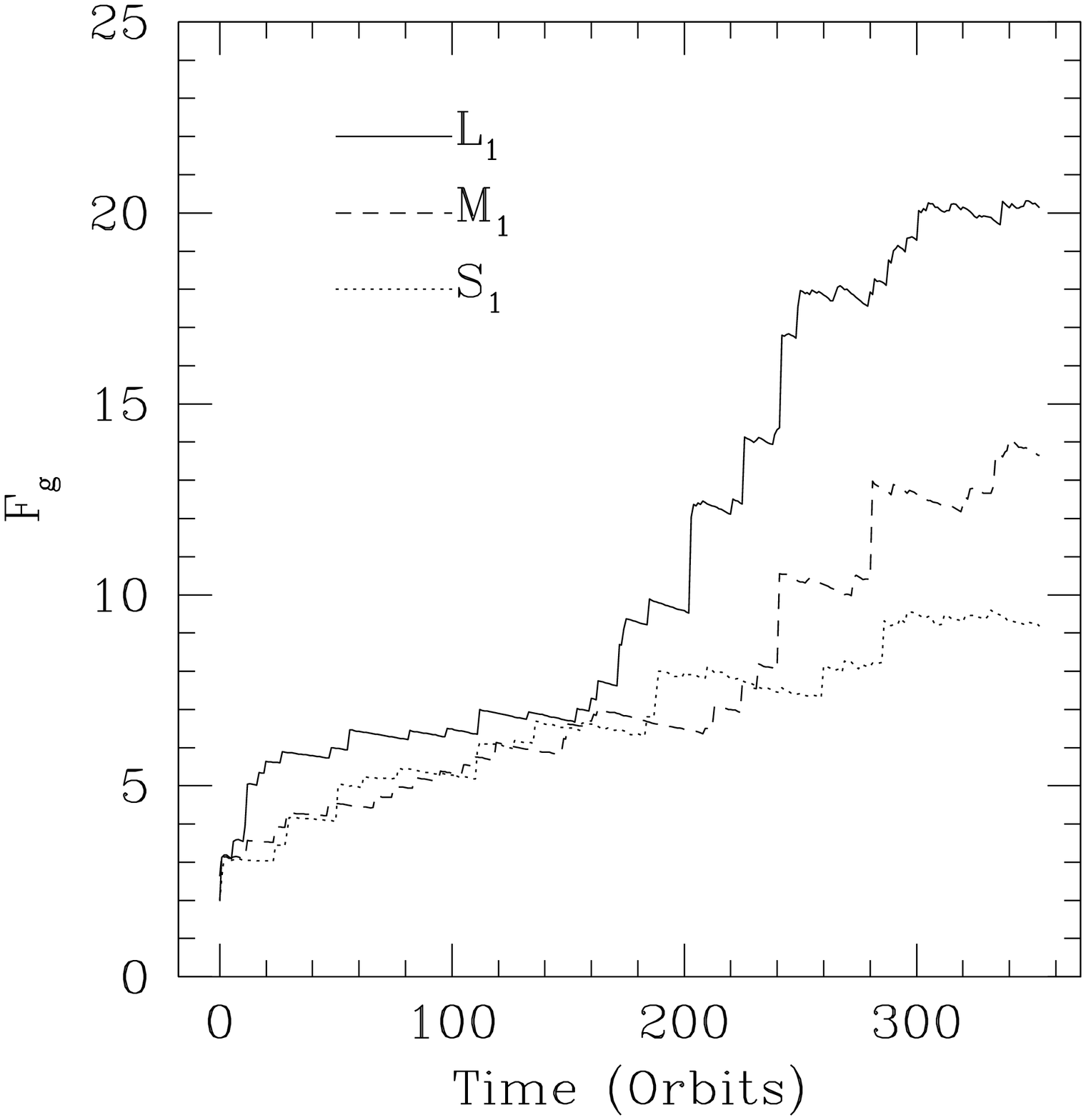}
\figcaption{\label{fig:eq-f_g} \small{Evolution of the gravitational focusing factor of the most massive particle in Simulations L$_1$ (solid line), M$_1$ (dashed line), and S$_1$ (dotted line). In larger simulations the focusing grows larger because the biggest particles are more massive, but the velocity dispersions are about equal in the three simulations (see Table 2).}}
\medskip

We evaluate the ratios of $e_{RMS}$ to sin $i_{RMS}$ in Fig.\
\ref{fig:eq-eiratio}. (Note the inclinations are in a regime such that the difference between sin $i$ and $i$ is about 1 part in $10^{10}$.)  The ratios begin at 2.35 but drop to $\sim 2.15$, similar to previous
results (Greenzweig and Lissauer 1990; Kokubo and Ida 1996). 

\medskip
\epsfxsize=8truecm
\epsfbox{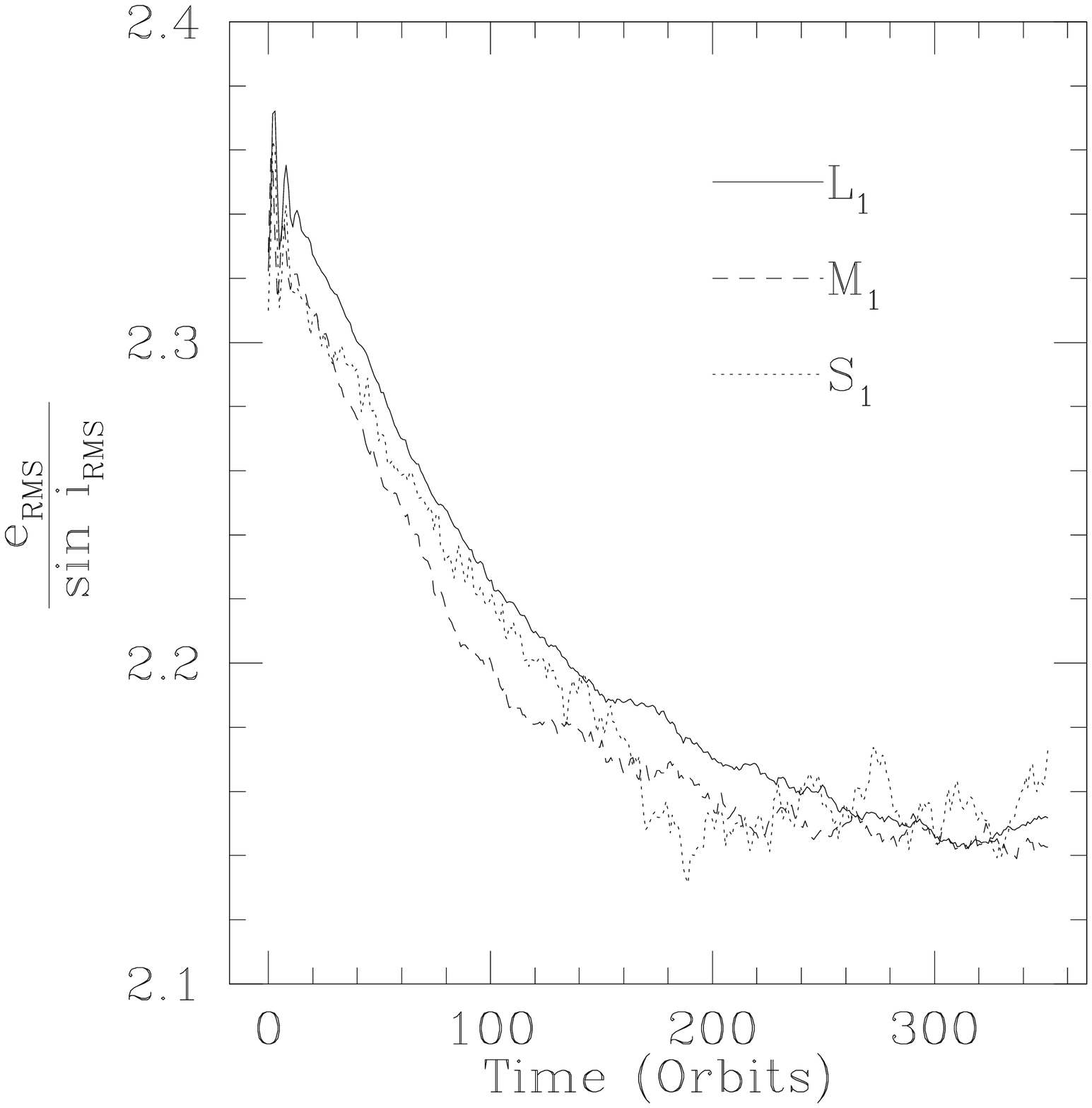}
\figcaption{\label{fig:eq-eiratio} \small{Evolution of $e_{RMS}/\sin i_{RMS}$ for Simulations L$_1$ (solid line), M$_1$ (dashed line), and S$_1$ (dotted line). The values drop from $\sim$2.35 to $\sim$2.15 over $\sim$200 orbits. At that time the ratios level out, which suggests the initial conditions were not in equilibrium.}}
\medskip

We investigate the role of dynamical friction by examining
$e_{RMS}$ and $i_{RMS}$ as a function of mass at $t_{1/2}$. For each populated bin,
we computed these values (if only one body occupied a bin we used its
$e$ and $i$ values), and show the results in Fig.\
\ref{fig:eq-dynfric}. As expected, larger masses are dynamically
colder than smaller masses, however they have greater energy, implying
that equipartition of energy is not achieved.

\medskip
\epsfxsize=8truecm
\epsfbox{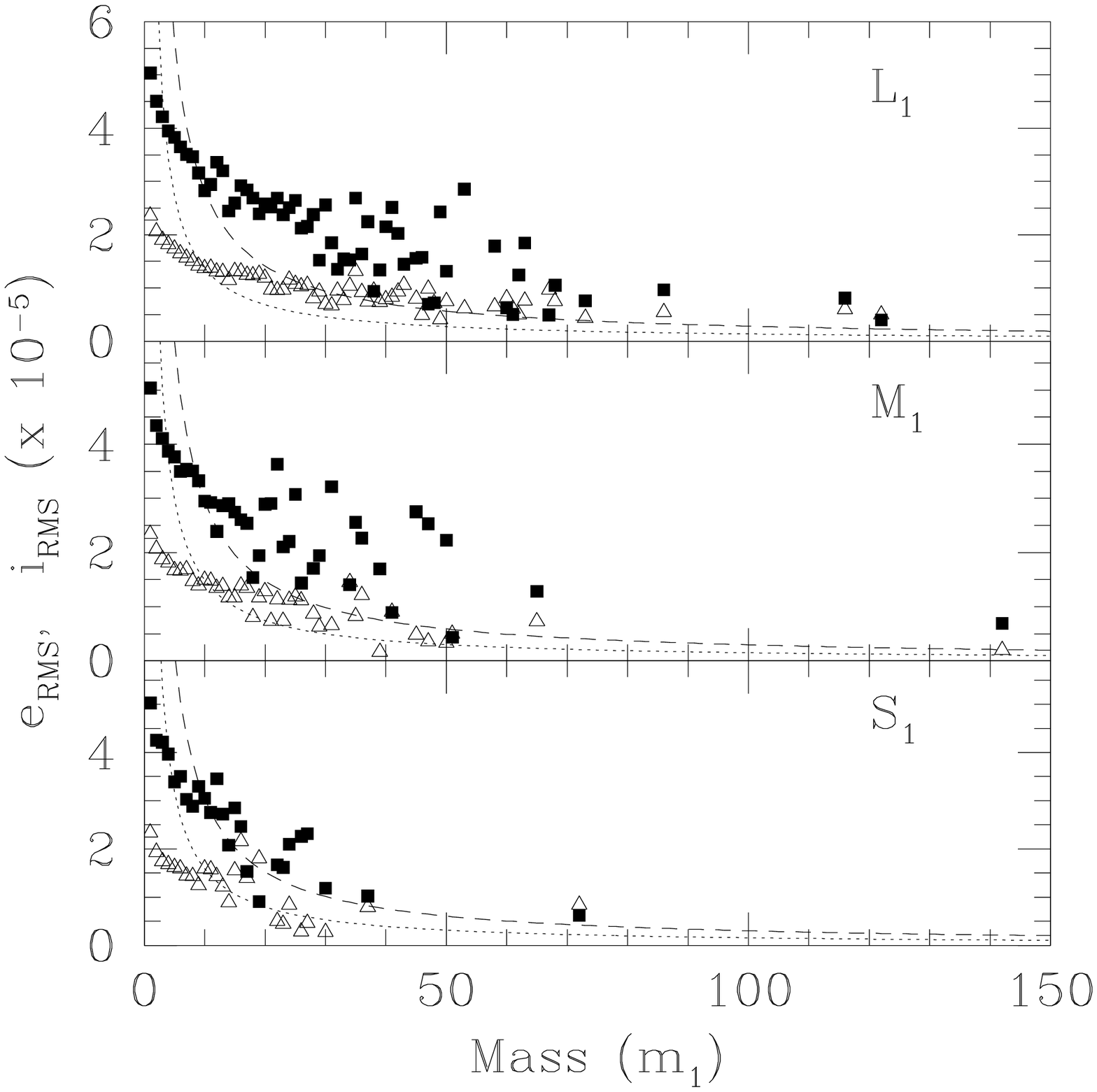}
\figcaption{\label{fig:eq-dynfric} \small{The values of $e_{RMS}$ and 
$i_{RMS}$ as a function of mass at $t_{1/2}$ for the three equilibrium 
simulations. Filled squares represent eccentricity, open triangles 
inclination. In the L$_1$ simulation, the largest mass planetesimal ($275 
m_1$) is 
not shown; its values are $e = 2.6 \times 10^{-6}$ and $i = 1.8 \times 
10^{-7}$. For 
reference, the dashed  line represents equipartition of energy in $e$, 
dotted in $i$, normalized to the values for $k = 10$.}} \medskip

\subsection{Rotation Rates}
We plot the rotation periods of all planetesimals in Simulation L$_1$ with $m > m_1$ at
$t_{1/2}$ in Fig.\ \ref{fig:eq-spin}. Only fourteen bodies had periods
longer than 1 day. Recall that initially all planetesimals have no
spin. We plot the spin distribution over
three mass ranges: $m = 2m_1$, $3m_1 \le m < 10m_1$, and $m \ge
10m_1$. The peak period is at 1.05 hours, which is less than the minimum period for a spherical gravitational aggregate of density 3 g/cm$^3$, see Eq.\ (\ref{eq:Pmin}). Thus, this distribution suggests that our assumption (which is the
standard one) of completely inelastic collisions resulting in mergers
overestimates planetesimal growth rates.

\medskip
\epsfxsize=8truecm
\epsfbox{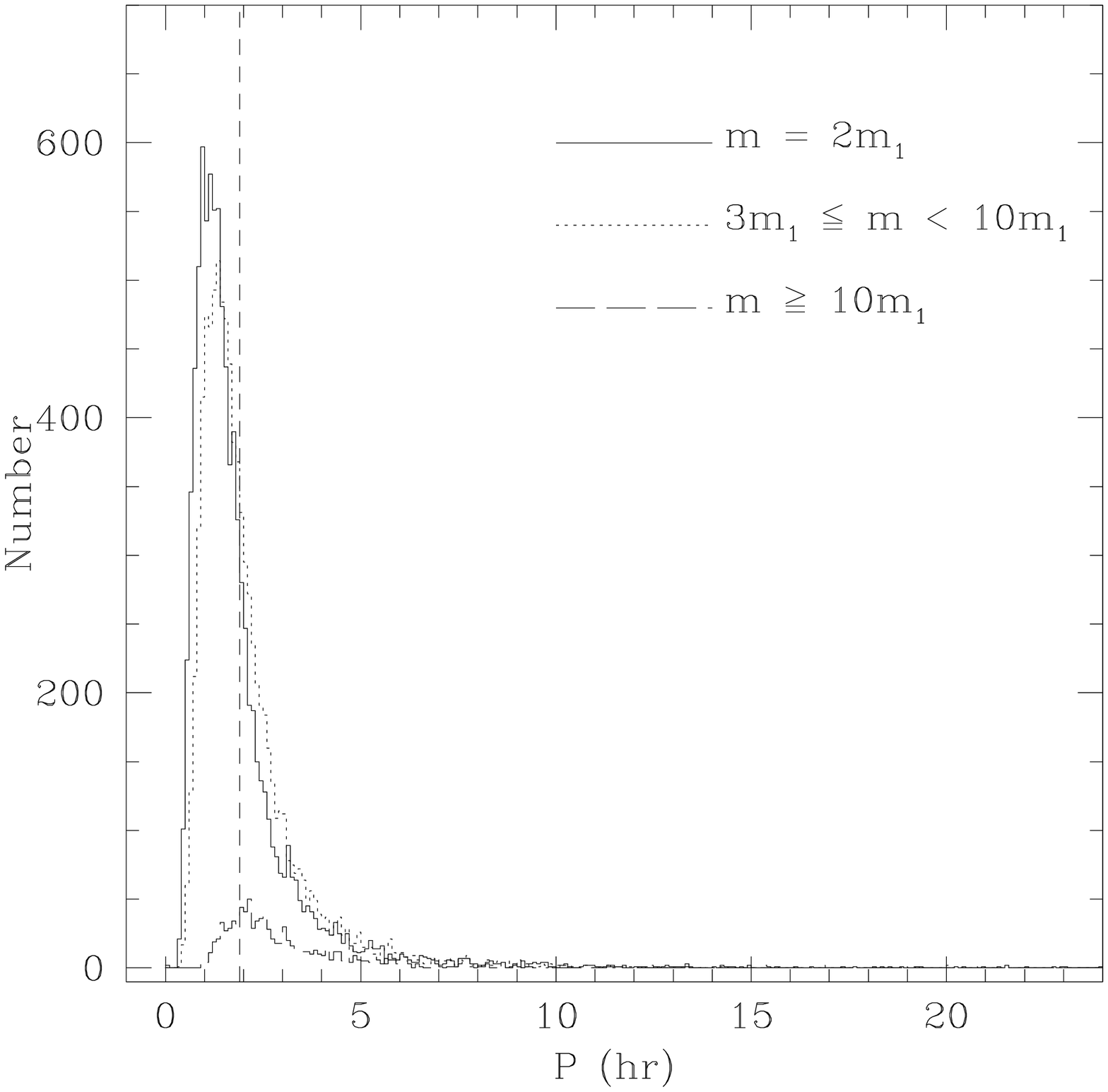}
\figcaption[]{\label{fig:eq-spin} \small{The final spin (at $t_{1/2}$) periods for the planetesimals in Simulation L$_1$. The dashed vertical line represents the minimum period of a gravitational aggregate.}}
\medskip

\subsection{Accuracy Tests}
In this subsection we quantify the accuracy of our results (see
$\S$2.5). $\S$4.4.1 measures the numerical accuracy of our code, and
$\S$4.4.2 describes the
statistical accuracy of our method.

\subsubsection{Numerical Accuracy}
In Fig.\ \ref{fig:eq-uw} we examine the validity of Simulation L$_1$ by
plotting $u$ and $w$. These parameters measure the center-of-mass
motion of the patch. Both $u$ and $w$ remain less 0.1 cm s$^{-1}$, about
1000 times smaller than the typical random velocities in the patch and
about $10^5$ times smaller than the shear across the patch, $\Omega W
= 4720$ cm s$^{-1}$. We therefore conclude that this
variation is tolerable (Wisdom and Tremaine 1988).

\medskip
\epsfxsize=8truecm
\epsfbox{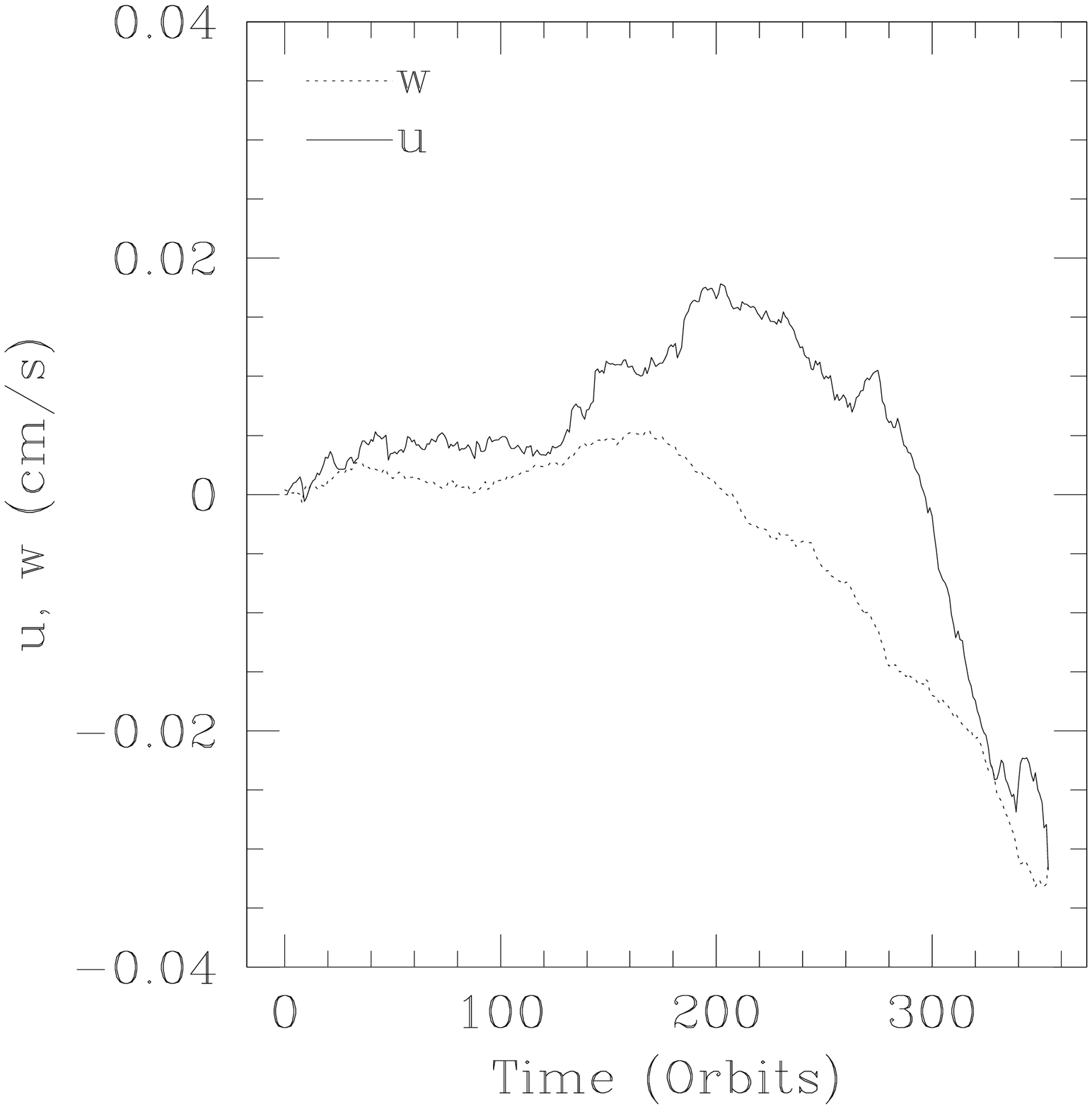}
\figcaption[]{\label{fig:eq-uw} \small{The evolution of the 
constants of motion for the nominal patch for Simulation L$_1$. The values of $u$ and $w$ 
vary at a level 3 orders of magnitude below that of the random motions.}}
\medskip

\subsubsection{Statistical Accuracy}
Although Simulation L$_1$ contains a large number of particles
(relative to modern $N$-body simulations), it still represents a very
small fraction of the terrestrial annulus. We therefore must
characterize the robustness of our results. The most critical
aspect of this experiment is the mass of the largest
particle. Should a particle reach a large enough mass that it
inappropriately dominates the dynamics of the patch, then our
assumptions have broken down.

First we consider $\beta$, Eq.\ (\ref{eq:beta}), which measures the
radial excursions of particles, see $\S$2.5. We find that after 354
orbits of Simulation L$_1$ that $\beta_{max-1}$ (the second largest
value) is still well below unity (Fig.\
\ref{fig:eq-bmax}). Therefore, the patch size for Simulation L$_1$ passes this requirement.

\medskip
\epsfxsize=8truecm
\epsfbox{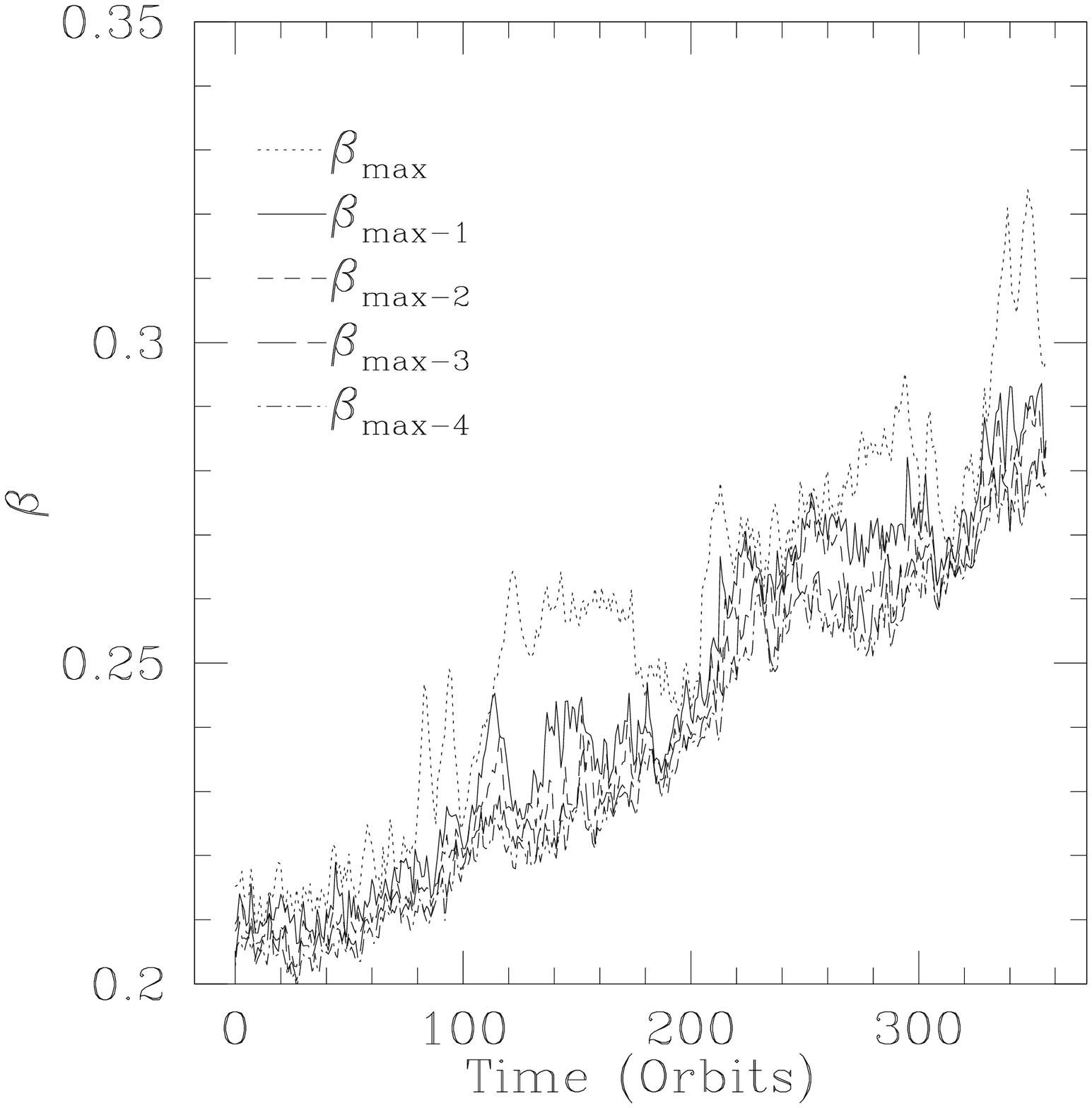}
\figcaption[]{\label{fig:eq-bmax} \small{The growth of the five largest values of $\beta$ in Simulation L$_1$ as a function of time. All values lie well below unity, indicating that no particle's epicycle is larger than the width of the patch. }}
\medskip

In Fig.\ \ref{fig:eq-stir}, we plot the evolution of the stirring
efficiency $S$ (Eq.\ [\ref{eq:stir}]) in the three baseline
simulations.  At $t_{1/2}$ the largest particle is about one-eighth
as effective at stirring as the rest of the swarm in Simulation L$_1$,
about one-seventh in M$_1$, and nearly one-fifth in S$_1$. These
values suggest we are nearing a situation in which the statistical
accuracy of this simulation cannot be confirmed, but such a situation
has not occurred yet.

\medskip
\epsfxsize=8truecm
\epsfbox{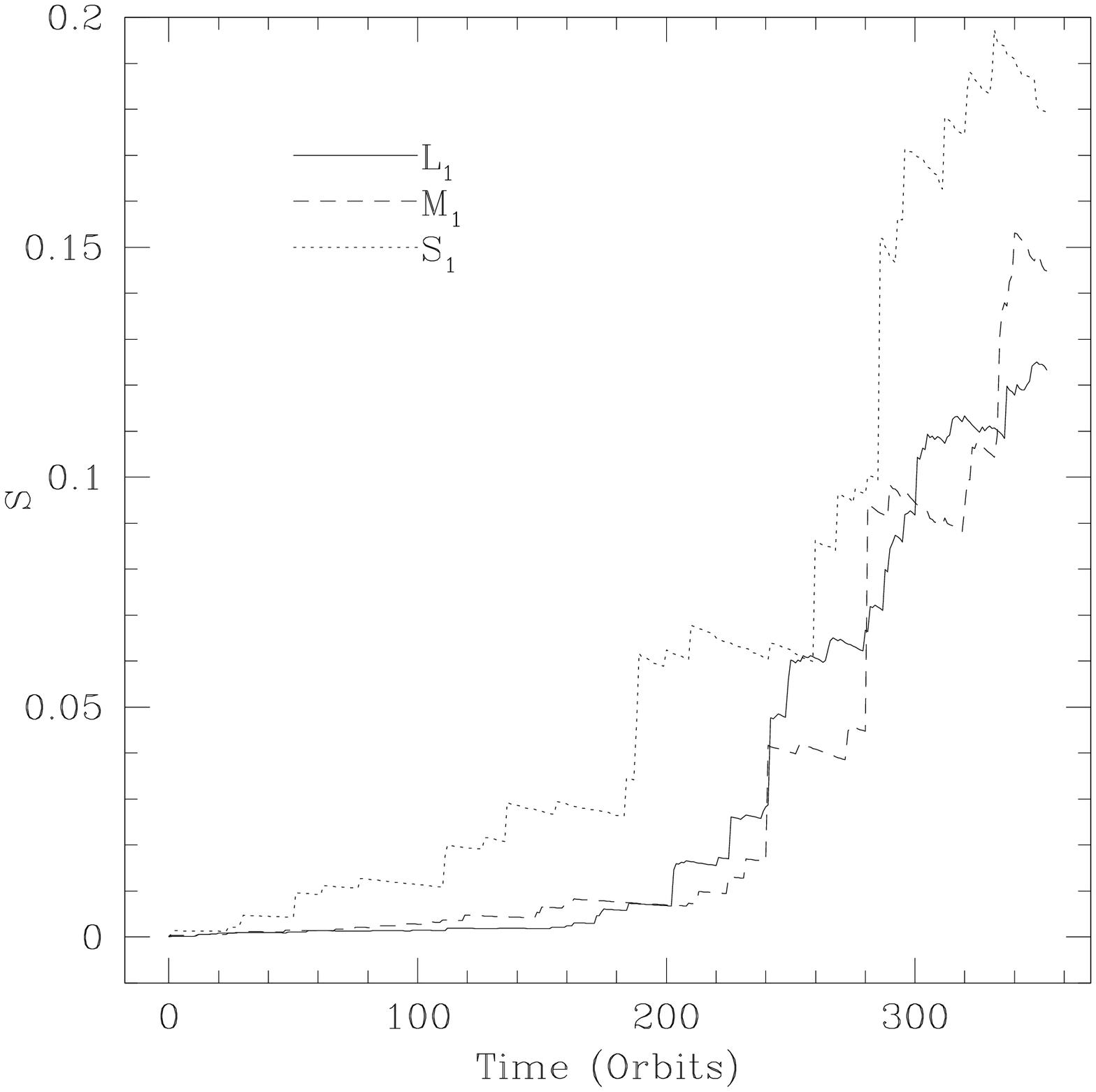}
\figcaption[]{\label{fig:eq-stir} \small{The stirring efficiency of the largest particles in Simulation L$_1$ (solid line), M$_1$ (dashed line), and S$_1$ (dotted line) relative to the rest of their respective swarms. }}
\medskip

Next we evaluate the distribution of $m_k^2N_k$ as a function of mass. This quantity measures how effective each mass bin is at stirring the patch. As
described in $\S$2.5, the distribution can reveal the possibility of
large-mass bodies beyond the boundaries of the patch that could
significantly change the dynamical character of the planetesimal
swarm. For this model, after about $t_{1/2}$ there is a significant
chance that large perturbers are close enough to affect the patches'
dynamics. In Fig.\ \ref{fig:eq-m2nll} we plot $m^2N_k$ vs.\ log $m$ at
four times. If $m^2N_k$ is decreasing as a function
of mass near the upper end of the mass distribution, then the patch is large enough to be a statistically
representative piece of the annulus, and the velocity distribution of
the patch should be close to that of the disk. If, alternatively, the
distribution is increasing, then there exist nearby large particles that
could dominate the stirring. We see at $t_{1/2}$ the distribution is
flat, suggesting we have reached the limit of our model. These curves suggest that results at later times may be inaccurate, hence our decision to exile those data in Appendix C. 

\medskip
\epsfxsize=8truecm
\epsfbox{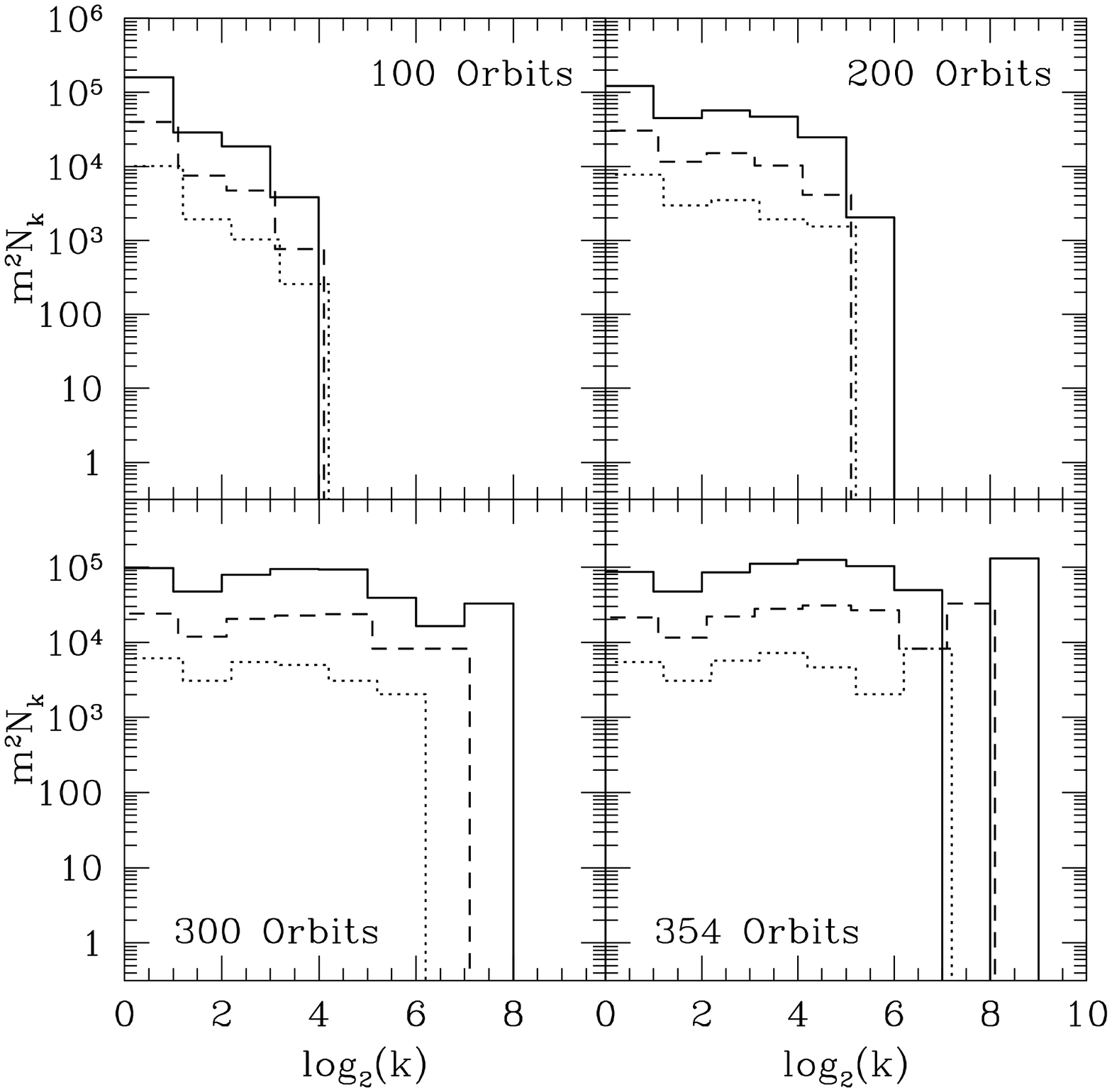}
\figcaption[]{\label{fig:eq-m2nll} \small{The stirring power of particles as a 
function of mass at 100 orbits (top left), 200 orbits (top right), 300 orbits (bottom left), and 354 orbits ($t_{1/2}$ for Simulation L$_1$; bottom right). The solid line represents Simulation L$_1$, dashed M$_1$, and dotted S$_1$. After 354 orbits, the slopes are roughly flat and shows that our patches might not be large enough to contain a statistically significant number of particles. Note that the Simulation M$_1$ data are offset by 0.1, and the S$_1$ data by 0.2.}}
\medskip

\section{Alternative Velocity Distributions}
In this section we describe the results of two simulations that begin
with different initial velocity dispersions than the baseline model. The
results of these simulations may be important since the initial
velocity dispersion of planetesimals is ill-constrained. Earlier growth may
occur too rapidly for the velocity dispersion to equilibrate ($v_{RMS} =
v_{esc}$). We thus explore a range of initial velocity dispersions. These
simulations also show how sensitive the results of $\S$4 are to
variations in the initial velocity dispersions. In Simulation
M$_{0.5}$ the initial velocity dispersion is set to $0.5v_{esc}$, in 
Simulation M$_2$ it is set to $2v_{esc}$.

For Simulations M$_{0.5}$ and M$_2$, $t_{1/2}$ occurred after 253 and
542 orbits, respectively. Figures \ref{fig:neq3-masslog} and
\ref{fig:neq3-c2f3} show the mass distributions at $t_{1/2}$ for
each of these models. As expected, when the velocity dispersion is
smaller, accretion proceeds faster, due to the increased gravitational
focusing. Moreover, the largest particles have a greater accretion advantage and so are more massive at $t_{1/2}$ in M$_{0.5}$ than in M$_2$.

\medskip
\epsfxsize=8truecm
\epsfbox{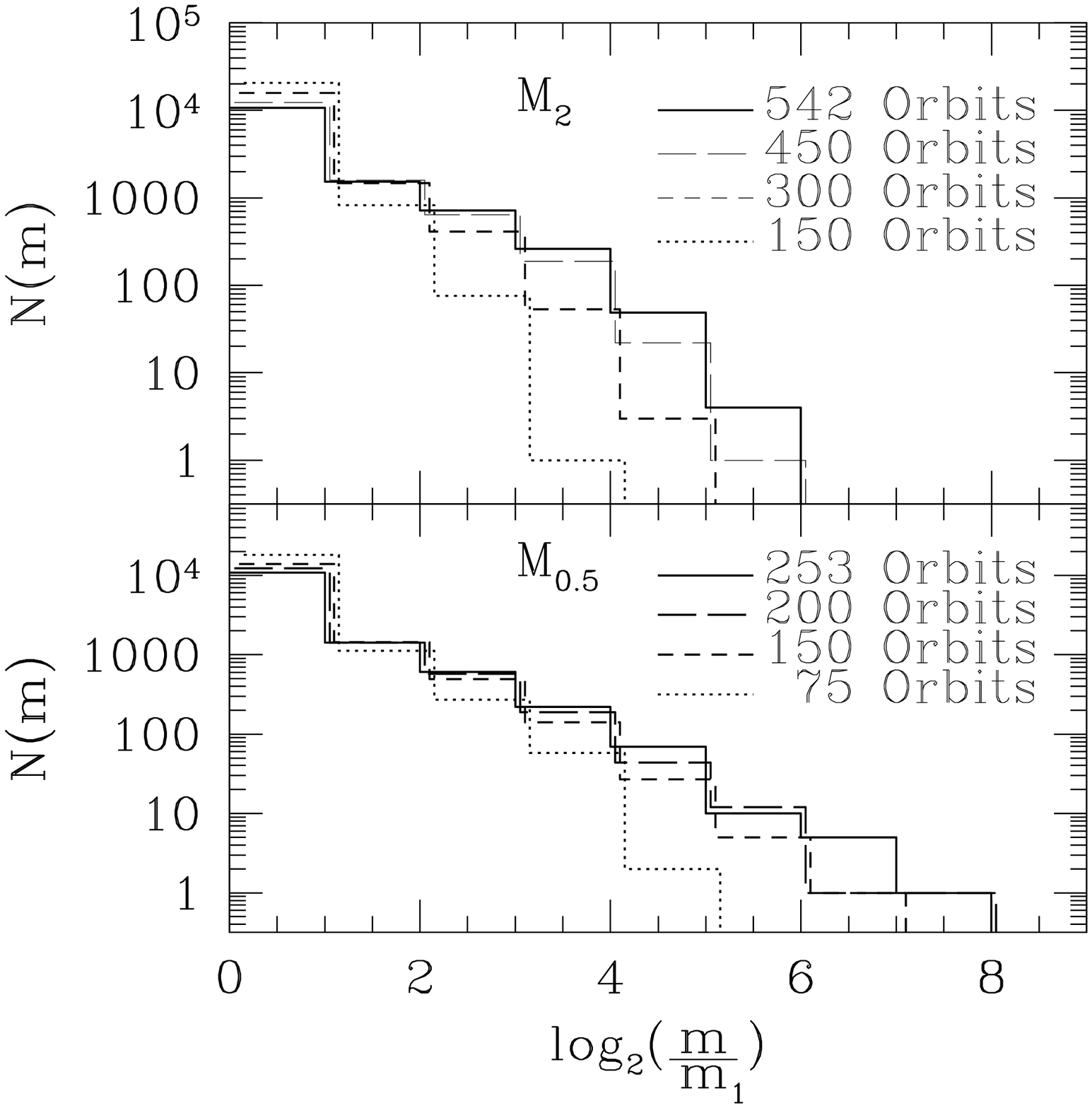}
\figcaption[]{\label{fig:neq3-masslog} \small{The mass functions of the 
non-equilibrium runs as a function of time. The earliest time is offset to the right by 0.05, the next time is offset by 0.1, etc. Larger initial velocity
dispersion suppresses runaway growth. These plots can be compared to Fig.\ \ref{fig:eq-mloglog}.}}
\medskip

In Fig.\ \ref{fig:neq3-c2f3} we show how our power-law fits (both with and without the $k = 1$ data) to the log-log mass
distribution, Eq.\ (\ref{eq:plaw}), compare to the
actual distributions at each Simulation's $t_{1/2}$. As with the
baseline models (see Fig.\ \ref{fig:eq-c2f3}), the 
exponential has a much larger unreduced $\chi^2$ value. The best fit parameters for these two
simulations at time $t_{1/2}$ are listed in Table 3.

\medskip
\epsfxsize=8truecm
\epsfbox{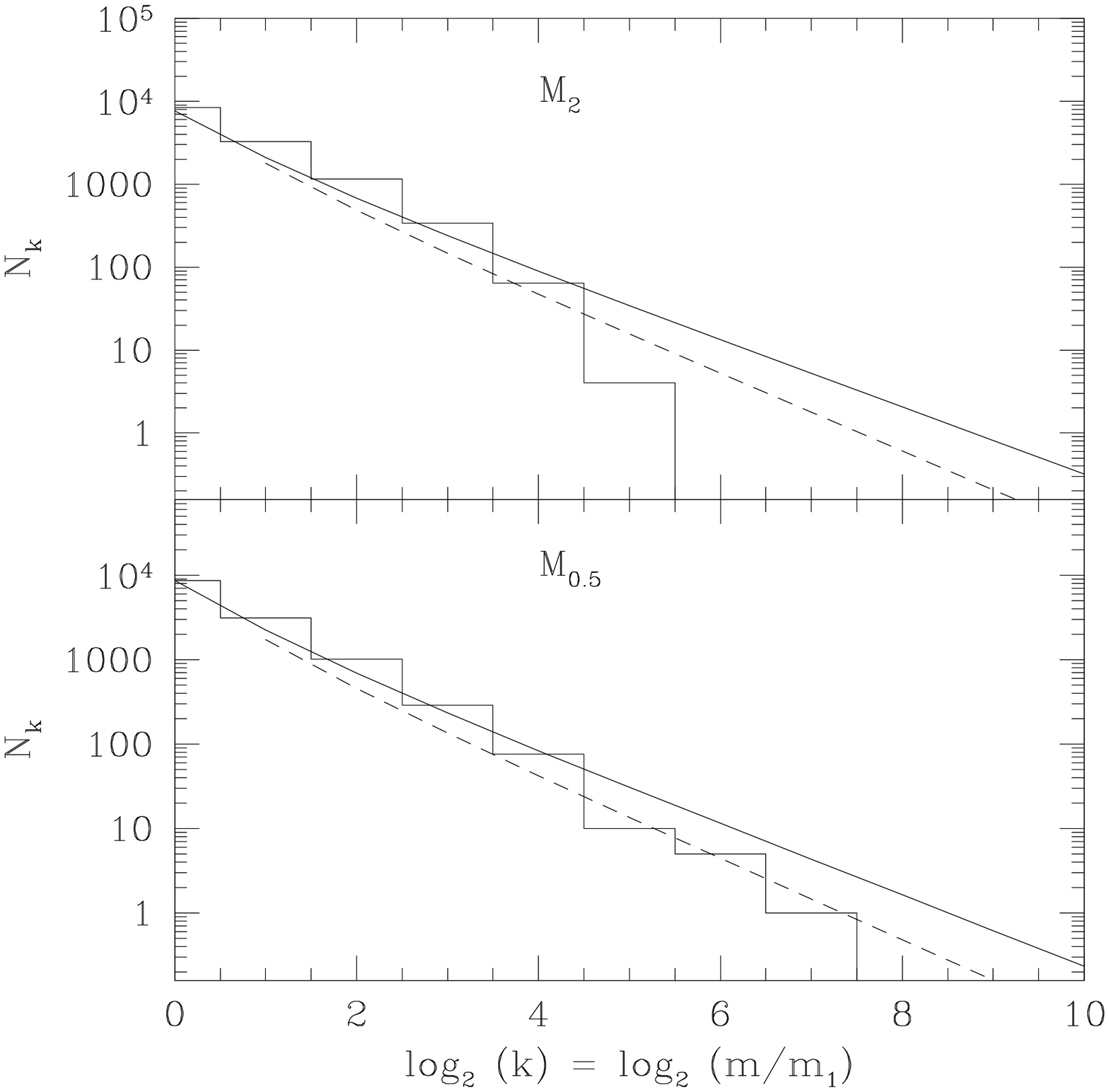}
\figcaption[]{\label{fig:neq3-c2f3} \small{Comparisons of the mass distributions at $t_{1/2}$ for the Simulations M$_2$ (top) and M$_{0.5}$ (bottom) and the best power law fits to the data, Eq. (\ref{eq:plaw}). The histograms are the mass distributions from the simulations, the straight lines are the fits (solid includes first bin, dashed does not).}}
\medskip

We continue by plotting the evolution of the largest mass particle for
each non-equilibrium run in Fig.\ \ref{fig:neq3-lm}. As in Fig.\
\ref{fig:eq-lm}, we also include the predictions of the Wetherill (1990) model. The linear solution is a good fit to the distribution of Simulation M$_2$, while the product solution appears to be a good fit to that of Simulation M$_{0.5}$.

\medskip
\epsfxsize=8truecm
\epsfbox{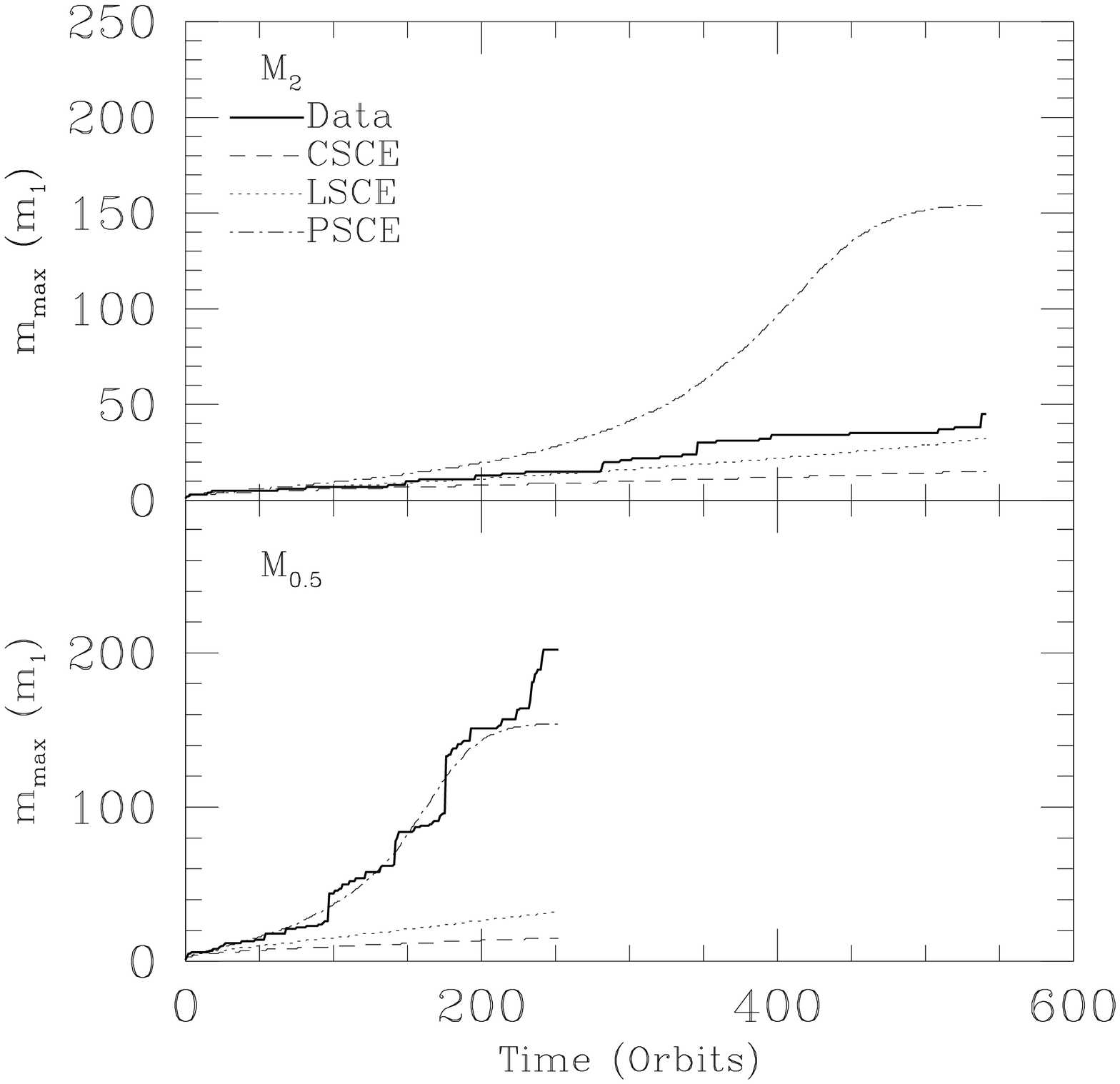}
\figcaption[]{\label{fig:neq3-lm} \small{The growth of the largest 
mass in each model as a function of time, and the predictions of the constant, linear, and product solution of the coagulation equation.}}
\medskip

In Fig.\ \ref{fig:neq2-vrms} we plot the evolution of the velocity
dispersions. In the top panel (Simulation M$_2$), we see that the
velocity dispersion actually drops initially, but then increases
similarly to the other runs. This decrease results from the inelastic nature of the collisions, as well as, to a lesser degree, the deposition
of translational kinetic energy into rotational kinetic
energy. However, we interpret this result with caution due to potential
inconsistencies in the perfect accretion model (see $\S$2.2). When
the initial velocity dispersion is too low, the dispersion quickly rises
(compare to Fig.\
\ref{fig:eq-vrms}). The final values of $F_g$ are 17.8 and 4.8 for Simulations M$_{0.5}$ and M$_2$, respectively. As with the baseline models, equipartition of energy has not occurred in these simulations.

\medskip
\epsfxsize=8truecm
\epsfbox{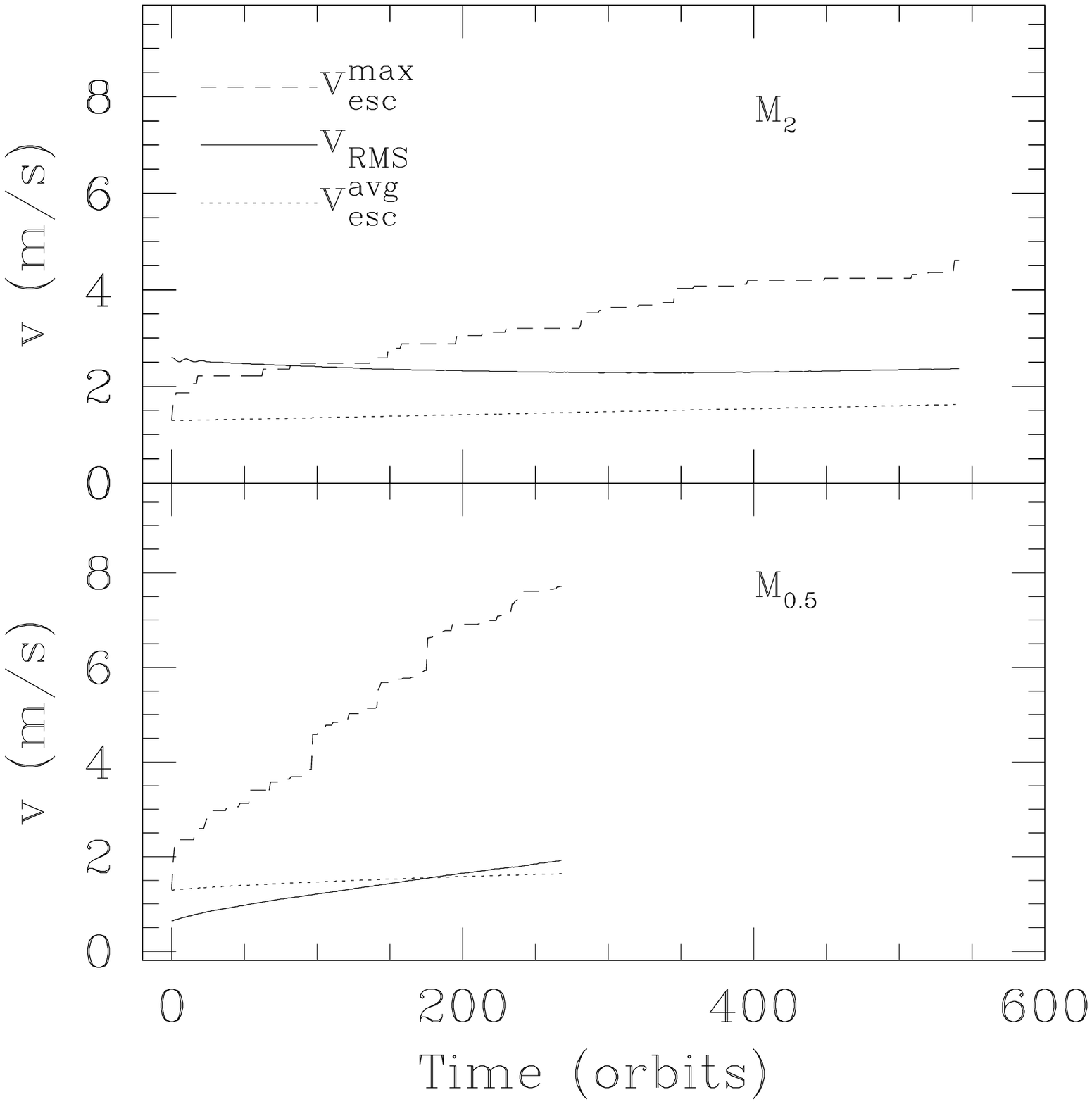}
\figcaption[]{\label{fig:neq2-vrms} \small{The evolution of the 
velocity dispersions in the non-equilibrium simulations. In Simulation M$_2$ (top panel), the dispersion actually drops toward equilibrium. The final values of $v_{RMS}$ are close to 2 m s$^{-1}$, as in the baseline case.}}
\medskip

In Fig.\ \ref{fig:neq3-spin} we plot the final spin distributions for
the non-equilibrium patches. The peaks lie below the minimum
gravitational aggregate period. Simulation M$_2$ consists of
especially fast rotators, due to a larger amount of kinetic energy of random motion available for transformation into rotational energy.

\medskip
\epsfxsize=8truecm
\epsfbox{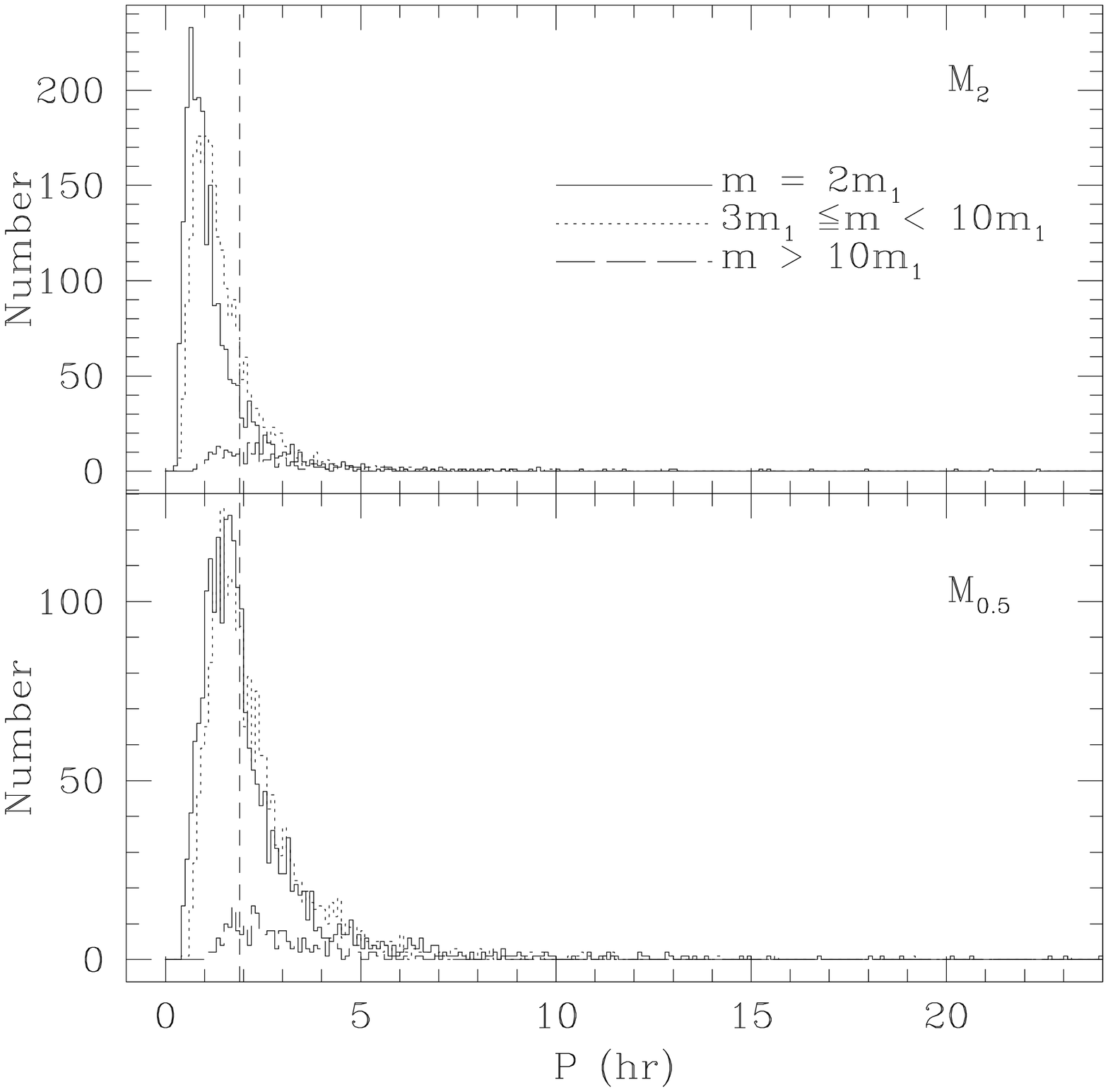}
\figcaption[]{\label{fig:neq3-spin} \small{The final spin distributions for merged particles for the non-equilibrium simulations. The vertical dashed line is the minimum period as defined by Eq.\ (\ref{eq:Pmin}). For Simulation M$_2$ very few particles have periods in excess of 10 hours, but Simulation M$_{0.5}$ has several long-period planetesimals.}}
\medskip

\section{Discussion}
Several results stand out in $\S\S$4 -- 5. First is that runaway
growth has not begun to occur for any particle in any simulation. Second, the
velocity dispersions of the patches remain close to the escape speed of the
average-mass particle. Third, the growth rate is moderately sensitive
to the initial velocity dispersion; changes of a factor of 2 in the
initial RMS velocity can excite or retard the early growth rate of the
most massive particles. Fourth, our collision model ($\S$2.2) is too simplified in that it assumes spherical particles experiencing perfect
accretion. Fifth, some aspects of the coagulation equation model the
growth well, but discrepancy of at least a factor of 2 are present,
and the best model (linear or product) appears to depend on the
initial value of $v_{RMS}$. Sixth, the power law fits represent a
realistic model of the actual mass distribution.

We present a summary of the results 
in Tables 2 -- 3. Comparing Simulation M$_1$ to the alternate-RMS velocity trial
s,
we see that its final properties lie between those of Simulations M$_{0.5}$ and
M$_2$. Therefore, $m_{max}$ and
$t_{1/2}$ depend upon with the initial velocity dispersion in a systematic manner
r.

Growth proceeds easily in all our models, despite no initial
seed. There is some indication from the final mass distributions that
the annulus will develop particles with a mass in excess of $10^4 m_1$
by $t_{1/2}$. By considering the stirring effects as a function of
mass we have found that by $t_{1/2}$ such large bodies could
significantly modify the dynamical properties of our
patches. Therefore, in order to continue our integrations further, we
must consider a larger patch, such that the curve of $S$ vs.\ $m$
turns over. Our simulations do not show how much larger the patch must
be, or indeed if any patch is adequate and $N$-body simulations at later
times must model a full annulus. Inconsistencies between our $N$-body
integration and the Wetherill (1990) model rule it out to
estimate the mass distribution. We note, however, that more
complicated models (\eg Kenyon and Bromley 2004) may make a better
match to our calculations, but the development and implementation of
such models was beyond the scope of this investigation. We encourage future
statistical researchers to use our results to verify their collision
kernels.

\medskip
\epsfxsize=8truecm
\epsfbox{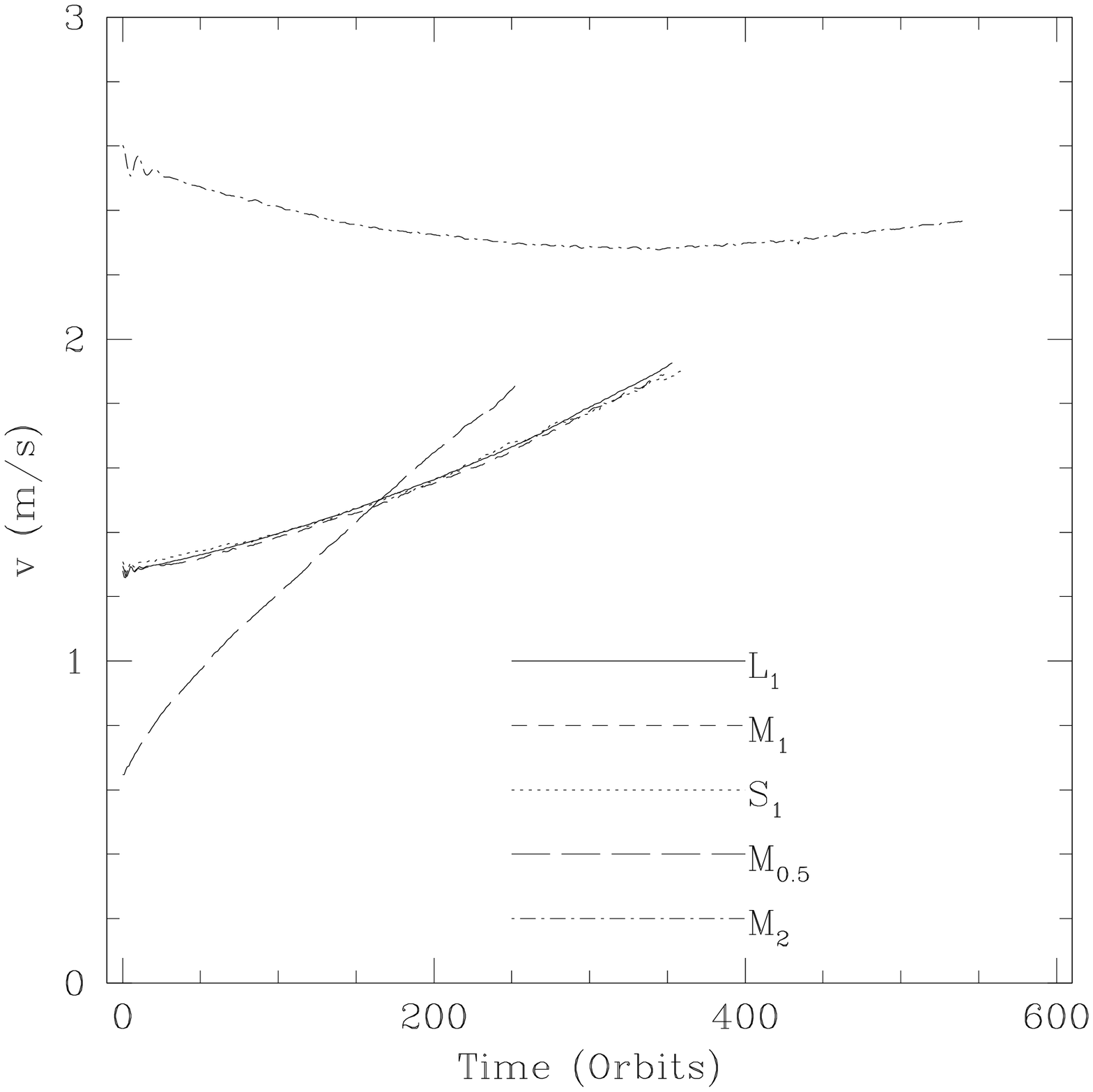}
\figcaption[]{\label{fig:vcomp}\small{Evolution of all RMS velocities.}} 
\medskip

In Fig.\ \ref{fig:vcomp} (see also Figs.\ \ref{fig:eq-vrms} and
\ref{fig:neq2-vrms}) we examine the different evolutions of the RMS
velocity dispersion for all our simulations. In each case, viscous stirring increases the velocity dispersion. Moreover, all have $v_{RMS} \approx 2$
m s$^{-1}$ at $t_{1/2}$. Note that at $t_{1/2}$ the average mass is $2m_1$,
which corresponds to an escape speed of 1.7 m s$^{-1}$. This equivalence
suggests that, at least early on, the velocity dispersion grows at
approximately the same rate as the escape speed of the typical mass
particle.

The perfect accretion model has produced several results that may be
spurious. The sum of all these issues, in our implementation, is that
orbital angular momentum is too easily transferred into rotational
angular momentum, and produces spin rates that are too high (see
Figs.\ \ref{fig:eq-spin} and
\ref{fig:neq3-spin}). Therefore the spin distributions presented
here should be not be regarded as physically realistic, and the mass distributions of
our simulations should be considered upper bounds.

Assuming the initial swarm of planetesimals is composed of gravitational 
aggregates, then we need larger simulations with more a realistic 
collisional model (see Leinhardt and Richardson 2005) in order to 
determine on the true nature of the post-collision particles. Presumably 
angular momentum is lost by shedding rubble from the surface. Three 
possibilities await this freed rubble: collision with other planetesimals, 
orbital migration via gas drag into the central star, or recollapse into 1 
km planetesimals via collapse (Goldreich \etal 2004). Given the number 
density of planetesimals at this stage of growth, the former seems the 
most likely. This limitation of our model demonstrates the need to perform 
similar simulations with a more realistic collisional model (\eg Leinhardt 
and Richardson 2005; Leinhardt \etal 2009).

Our analytic fits to the mass distributions show that exponential fits
do not match the data (see Table 3). However the power law does provide a reasonable fit for most of our
models. Removing the $k = 1$ 
bin from our fits results in a significant decrease in the unreduced
$\chi^2$ values, as shown in Table 3. We conclude
that runaway growth has not begun in our simulations.

All of our simulations show that growth from 1 km planetesimals can
proceed quickly. In fact, it proceeds so rapidly that our model breaks
down in just a few hundred orbits (see $\S$4.4.2, Fig.\
\ref{fig:eq-m2nll}). Therefore the only way to realistically proceed
beyond $t_{1/2}$ is to enlarge our patches. As we expand our patches,
the number of particles increases. Given that Simulation L$_1$
required $\sim 30,000$ node hours to complete on the Columbia
Supercomputer at NASA Ames, and computation time scales as $N$log $N$,
we may not be able to expand our patches such that they are both
statistically accurate and computationally tractable. Our results
therefore suggest that the patch model may be inadequate to model
later stages of terrestrial planet formation.

\section{Conclusions}
We have performed the first $N$-body simulations of growth from 1 km
planetesimals. The initial conditions of our runs were chosen to be
similar to those believed to have existed in our protosolar disk;
substantially different parameters may be appropriate for the initial
stages of growth in extreme exoplanetary systems (Lissauer \&
Slartibartfast 2008). These simulations required hundreds of thousands
of node hours of supercomputer time, using an advanced $N$-body code
designed specifically to examine systems with large $N$. Although
numerous shortcuts and approximations were incorporated in our model,
we believe that our results provide insights into planetesimal growth
and lay a foundation for future investigations.

Planetesimal growth from a uniform swarm of 1 km-sized planetesimals
proceeds in a stochastic fashion. Our results have confirmed some of
the trends seen in the semi-analytical research (Greenberg \etal 1978;
Wetherill 1990; Weidenschilling \etal 1997) into the growth of 1 km
planetesimals. Until more realistic models of fragmentation can be
implemented in $N$-body code, statistical methods are the only 
feasible
approach to address fragmentation.

Some of the assumptions of our model broke down relatively quickly, 
demonstrating the limits of the patch approximation in modeling planetary 
accretion. Nonetheless, our results suggest new directions of research for 
this epoch of planet formation. In particular, a more realistic 
collisional model (one in which additional small particles carry away 
excess angular momentum, \ie fragmentation) seems most important. Such a 
model may suppress growth (as well as eliminating unphysical spins), and 
hence planetesimals would not grow so quickly. However, this approach is 
considerably more complex and numerically intensive than those presented 
here. Until these modifications can be made, our results represent the 
most accurate model of 1 km planetesimal growth available.

\section{Acknowledgments}
We thank Richard Greenberg, Glen Stewart, Don Brownlee, Guillermo
Gonzalez, Paul Hodge, and Brian Jackson for useful discussions, and an
anonymous referee for valuable suggestions. This research was funded
primarily by NASA's Terrestrial Planet Finder Foundation Science/Solar
System Origins under grant 811073.02.07.01.15. Simulations were
performed on the Lemieux supercomputer at the Pittsburgh Supercomputer
Center and the Columbia supercomputer at NASA Ames Research
Center. Computers used to analyze these simulations were provided by
the University of Washington's Student Technology Fund. Parallel code
development was partly supported by NSF grant PHY-0205413. Rory Barnes
also thanks KFB for his unwavering encouragement, and acknowledges 
additional support from NASA's Planetary Geology and
Geophysics grant NNG05GH65G and NASA's Graduate Student
Research Program. Thomas Quinn was partly supported by the NASA
Astrobiology Institute. Jack J. Lissauer acknowledges support from
NASA's PG\&G program, grant 811073.02.01.01.12. Derek
C. Richardson acknowledges support from NASA Grant NAG511722
issued through the Office of Space Science.

\section*{APPENDIX A: List of Symbols and Abbreviations}
\noindent $A = $ arbitrary constant\\
$A_{lj} = $ collisional probability coefficient in coagulation theory\\
$a = $ semi-major axis\\
$B = $ arbitrary constant\\
$b = $ power law fit parameter\\
$b' = $ power law fit parameter with $m = m_1$ bin excluded\\
$C = $ arbitrary constant\\
$c = $ exponential fit parameter\\
$c' = $ exponential fit parameter with $m = m_1$ bin excluded\\
$D = $ arbitrary constant\\
$e = $ eccentricity\\
$e_{RMS} = $ root mean square eccentricity of particles in a patch\\
$F_g = $ gravitational focusing factor\\
$f = $ fraction of particles relative to initial number\\
$G = $ Newton's gravitational constant\\
$i = $ inclination\\
$i_{RMS} = $ root mean square inclination of particles in a patch\\
$j = $ counter in coagulation equations\\
$k = $ ratio of particle mass to mass of 1 km planetesimal\\
$k_R = $ mass of a runaway particle relative to a 1 km planetesimal\\
L$_1$ = Largest baseline simulation\\
$l = $ counter in coagulation equations\\
M$_{0.5}$ = simulation with the initial velocity dispersion magnitude set to half that of Simulation L$_1$\\
M$_1$ = medium-sized baseline simulation\\
M$_2$ = simulation with the initial velocity dispersion magnitude set to twice that of Simulation L$_1$\\
$M_{ann} = $ mass in an annulus of the protoplanetary disk\\
$M_{patch} = $ total mass inside a patch\\
M$_\odot = $ mass of the Sun\\ 
M$_\oplus = $ mass of the Earth\\
$m = $ mass\\
$m_1 = $ mass of 1 km planetesimal\\
$m_{crit} = $ particle mass at which, in one orbit, it collides with an equal mass of gas\\
$m_{max} = $ largest mass in a simulation\\
$m_{max-1} = $ second largest mass in a simulation\\
$m_{max-4} = $ fifth largest mass in a simulation\\
$m_{max-9} = $ tenth largest mass in a simulation\\
$m_{pl}= $ mass of planetesimal\\
$<m> = $ average mass of planetesimals\\
$N = $ number of bodies in a simulation\\
$N_0 = $ initial number of planetesimals in a simulation\\
$N_{patch} = $ number of bodies in a patch\\
$N_k = $ number of bodies in a mass bin\\
$n_k = $ number density of particles in mass bin $k$\\
$P = $ orbital period\\
$P_{peak}$ = peak of the spin period distribution\\
S$_1$ = Simulation with same initial properties as L$_1$, but only 1/16 the size\\
PIAB = Particle-in-a-box\\
$R = $ planetesimal radius\\
$R_{max} = $ radius of largest planetesimal\\
$r = $ heliocentric radius\\
$r_{patch} = $ heliocentric radius of the center of a patch\\
$t = $ time\\
$S = $ stirring power of largest mass relative to that of all other bodies\\
S$_1$ = smallest-sized baseline simulation\\
$t = $ time\\
$t_{1/2} = $ the time required to reduce the total number of particle at distance $r$ by 2\\
$t_{base} = $ longest timestep in a simulation\\
$t_{cross} = $ crossing time for two planetesimals\\
$t_{min} = $ minimum timestep in a simulation\\
$u = $ center-of-mass speed of a patch in the $x$-direction\\
$V = $ volume\\
$v = $ velocity\\
$v_{esc} = $ escape speed of a planetesimal\\
$v_{RMS} = $ root mean square speed of a patch\\
$v_x = $ speed in $x$-direction\\
$v_y = $ speed in $y$-direction\\
$v_z = $ speed in $z$-direction\\
$W = $ size scale of a patch\\
$w = $ center-of-mass speed of a patch in the $y$-direction\\
$x = $ Cartesian coordinate that mimics heliocentric distance\\
$x_g = x$ position of guiding center of a planetesimal's epicycle\\
$y = $ Cartesian coordinate that mimics azimuthal position\\
$y_g = y$ position of guiding center of a planetesimal's epicycle\\
$z = $ height above/below midplane\\
$Z_0 = $ scale height of planetesimal disk\\
$\beta = $ radial excursions of a planetesimal due to eccentricity\\
$\beta_{max} = $ largest radial excursion of a particle in a patch\\
$\beta_{max-1} = $ second largest radial excursion of a particle\\
$\beta_{max-2} = $ third largest radial excursion of a particle\\
$\beta_{max-3} = $ fourth largest radial excursion of a particle\\
$\beta_{max-4} = $ fifth largest radial excursion of a particle\\
$\zeta = $ number of rungs in a simulation\\ 
$\eta = $ scale factor to determine timesteps\\
$\Theta = $ maximum apparent size of a cell for PKDGRAV to only use the hexadecapole moment\\
$\theta = $ azimuthal position of a planetesimal in heliocentric coordinates\\
$\nu_1 = $ collisional probability coefficient in constant solution of coagulation equation\\
$\nu_2 = $ collisional probability coefficient in linear solution of coagulation equation\\
$\nu_3 = $ collisional probability coefficient in product solution of coagulation equation\\
$\rho = $ volume mass density\\
$\rho_0 = $ volume mass density at midplane\\
$\rho_{pl} = $ mass density of a planetesimal\\
$\Sigma = $ surface density\\
$\Sigma_0 = $ coefficient that scales surface density of planetesimal disk\\
$\sigma = $ physical cross-section of a planetesimal\\
$\sigma_{pl} = $ gravitationally enhanced cross-section of a planetesimal\\
$<\sigma_{pl}> = $ average gravitationally enhanced cross-section of planetesimals\\
$\tau = $ mean free time between planetesimal physical collisions\\
$\phi = $ gravitational potential\\
$\chi^2_b = $ unreduced $\chi^2$ value for power law fit to mass distribution\\
$\chi^2_{b'} = $ unreduced $\chi^2$ value for power law fit to mass distribution with $m = m_1$ mass bin excluded\\
$\chi^2_c = $ unreduced $\chi^2$ value for exponential fit to mass distribution\\
$\chi^2_{c'} = $ unreduced $\chi^2$ value for exponential fit to mass distribution with $m = m_1$ mass bin excluded\\
$\Omega_{patch} = $ Keplerian orbital frequency of a patch\\
$\Omega_z = $ vertical frequency due to Keplerian motion and the mass of the disk\\

\section*{APPENDIX B: The Coagulation Equation} 
\renewcommand{\theequation}{B\arabic{equation}}
% redefine the command that creates the equation no.
\setcounter{equation}{0}  % reset counter 

Here we summarize the basics of the coagulation equation as presented by Wetherill (1990). The discrete form of the coagulation equation is
\begin{equation}
\label{eq:coag}
\frac{dN_k}{dt} = \frac{1}{2}\sum_{l+j=k} A_{lj}n_ln_j - n_k\sum_{l=1}^{\infty}A_{lk}n_l
\end{equation}
where the generic indices $j$, $k$ and $l$ are just the ratio $m/m_1$.
In Eq.\ (\ref{eq:coag}), $N_k$ is the number of particles in bin $k$,
$A_{jl}$ is the probability of collision, $n_j$ is the total number
density of particles of mass $j$, $m$ is the mass, and $m_1$ is the
mass of a 1 km planetesimal. The first term is the probability of all
combinations of particles of mass $l$ and $j$ that sum to equal the current
mass bin $k$. The mass bin is the quantum of the mass spectrum. The
factor of 1/2 prevents the summation from counting all collisions twice (when $l = j$). The second term is the loss of particles from mass bin $k$ to
larger mass bins. Note that, despite the discrete nature of Eq.\ (\ref{eq:coag}), for $t > 0$, it can predict a fractional number of particles in each bin.

The collisional probability coefficient, $A_{lj}$, is a
function of the relative velocity of particle $l$ to $j$, their
masses, the number density of each bin, and the volume being considered. Therefore
\begin{equation}
\label{eq:Adef}
A_{lj} = A_{lj}(v_{rel},m_l,m_j,n_l,n_j),
\end{equation}
where $n_{l}$ ($n_j$) is the number density of particles with mass $l$ ($j$).
Three forms of this function have been examined. The simplest solution
is
\begin{equation}
\label{eq:Aconst}
A_{lj} = \nu_1,
\end{equation}
a constant. For linear dependence we assume
\begin{equation}
\label{eq:Alin}
A_{lj} = \nu_2 (l+j),
\end{equation}
a constant times the sum of the masses, and the dependence
on velocities and densities has been subsumed into $\nu_2$. These two
possible solutions both fall under the category of orderly growth. A third solution, which is
proportional to the product of the masses, assumes
\begin{equation}
\label{eq:Aprod}
A_{lj}= \nu_3 lj.
\end{equation}
At any given orbit, Wetherill also gives equations for the collisional probabilities as a function of time:
\begin{equation}
\label{eq:nu1}
\nu_1 = \frac{2(1-f)}{N_0ft},
\end{equation}
\begin{equation}
\label{eq:nu2}
\nu_2 = -\frac{\textrm{log} f}{N_0t},
\end{equation}
and
\begin{equation}
\label{eq:nu3}
\nu_3 = \frac{2(1-f)}{N_0t}.
\end{equation}

The solutions to the constant and sum forms are
\begin{equation}
\label{eq:constcoag}
N_k = N_0f^2(1-f)^{k-1},
\end{equation}
and
\begin{equation}
\label{eq:lincoag}
N_k = N_0\frac{k^{k-1}}{k!}f(1-f)^{k-1}e^{-k(1-f)},
\end{equation}
respectively, where $N_0$ is the initial number of particles, and$f$
is just the fraction of the number of particles remaining at time $t$,
\begin{equation}
\label{eq:fdef}
f = \frac{N_{tot}(t)}{N_0} = \frac{\sum_{l=0}^{\infty}N_l}{N_0}.
\end{equation}
The product solution to the coagulation equation is
\begin{equation}
\label{eq:prodcoag}
N_k = N_0\frac{(2k)^{k-1}}{k!k}(1-f)^{k-1}e^{-2k(1-f)},
\end{equation}
and yields runaway growth. This however leads to the natural problem that a runaway particle is a
special particle, and it should not be treated as typical. This marks
the breakdown in the PIAB model. These solutions to the coagulation equation are used in $\S$4. In the text we refer to Eq.\ (\ref{eq:constcoag}) as constant coagulation, Eq.\ (\ref{eq:lincoag}) as linear coagulation, Eq.\ (\ref{eq:prodcoag}) as product coagulation.

\section*{Appendix C: The Baseline Model to 2000 Orbits}
In the spirit of Icarus' flight to the Sun, we present results for the baseline simulations from
$t_{1/2}$ to 2000 orbits here. As shown in $\S$4.4.2, after $t_{1/2}$
larger mass bodies may significantly alter the velocity dispersion in the
patches. However, the locations of such particles relative to the
patch are unknown. The synodic period across the radial width of the
patch is about 2400 orbits. We may therefore presume that by 2000
orbits, a large mass has entered the patch and significantly altered
the dynamics, but the time and magnitude of the changes are
unknown. Although the results in this appendix suffer from significant
inconsistencies, we nonetheless present them here, as they
represent the only $N$-body simulation of growth from 1 km
planetesimals to date. Table 4 lists some of the properties of the
baseline models at 2000 orbits. \textit{Results in this appendix
should not be regarded as physically realistic simulations of
planetesimal growth beyond $t_{1/2}$!}

Figure \ref{fig:app-mloglog} presents the mass distribution of
particles in a format similar to Fig.\ \ref{fig:eq-mloglog}. At 2000
orbits, three particles in L$_1$ and one in M$_1$ have reached masses
larger than 5000 $m_1$. By comparison, the fourth largest planetesimal
in L$_1$ is one-sixth as massive, 814 $m_1$, see Fig.\
\ref{fig:app-lm}.

\medskip
\epsfxsize=8truecm
\epsfbox{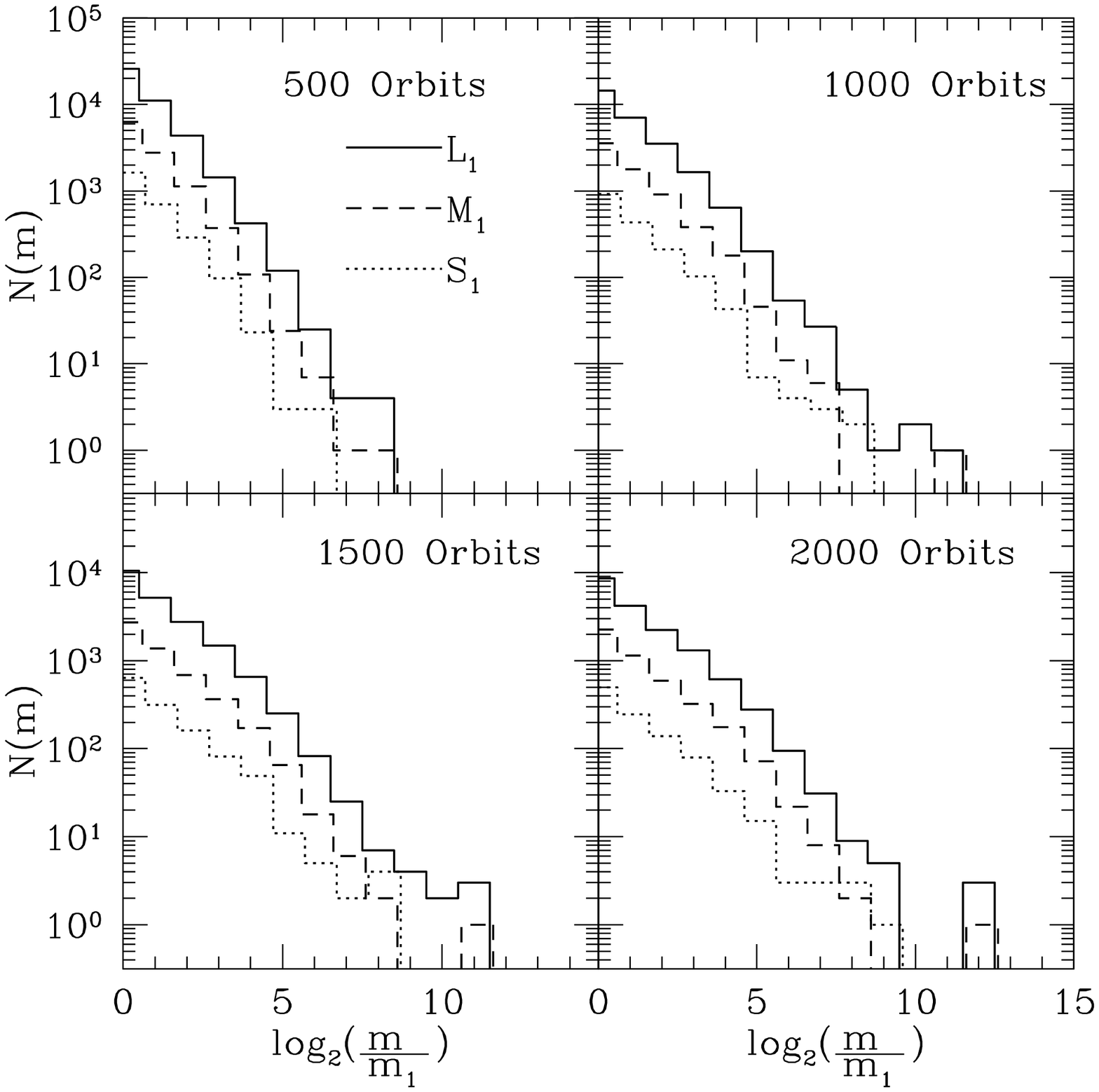}
\figcaption[]{\label{fig:app-mloglog} \small{Mass spectrum at various times during the evolution of the baseline patches in log-log format. Note that the M$_1$ simulation has been offset by 0.1 and S$_1$ by 0.2.}}
\medskip

Figure \ref{fig:app-c2f3} presents the mass functions at 2000 orbits,
along with the associated analytic fits to the data. At 2000
orbits the power law has become a better fit to the data than at
$t_{1/2}$. Table 5 lists the fit parameters (and corresponding measures
of goodness of fit) at 2000 orbits. The values of the key parameters $b$ and $b'$ all cluster near 1.9. At 2000 orbits the
exclusion of the $k = 1$ bin does not cause dramatic changes in the
the unreduced $\chi^2$ values, which is not surprising since only 10\% of the particles remain at $k = 1$.

\medskip
\epsfxsize=8truecm
\epsfbox{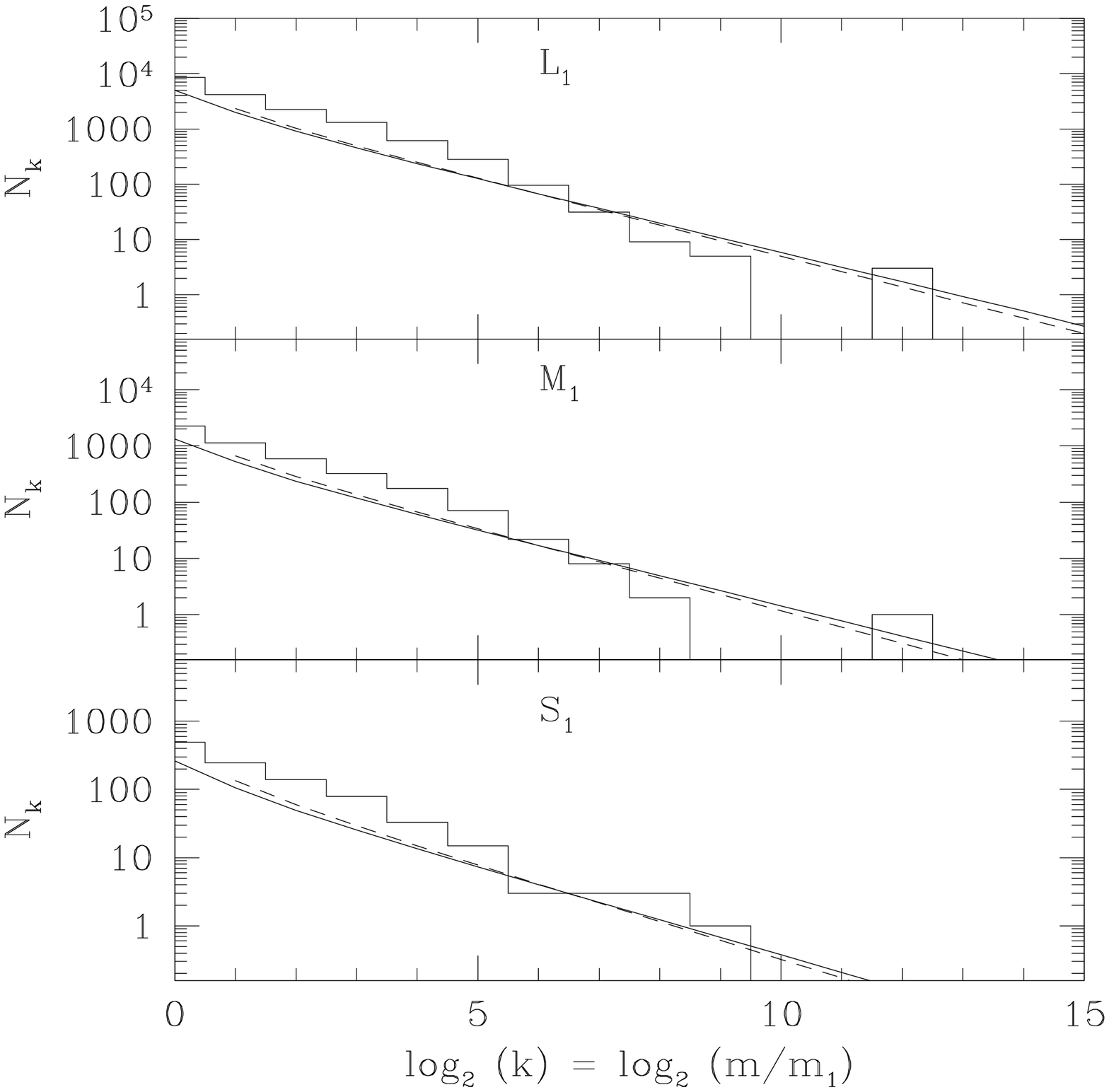}
\figcaption[]{\label{fig:app-c2f3} \small{The mass function of Simulations 
L$_1$ (top), M$_1$ (middle) and S$_1$ (bottom) and the associated power-law fits to the data; see Eq. (\ref{eq:plaw}) and Table 5. }}
\medskip

Figure \ref{fig:app-lm} shows the evolution of some of the largest
particles in the L$_1$ patch. At 713 orbits, the two largest particles
in the patch merge to form a 1610 $m_1$ mass object. This merger has
important consequences for our assumptions about the statistical
accuracy of our patch, as shown below.

\medskip
\epsfxsize=8truecm
\epsfbox{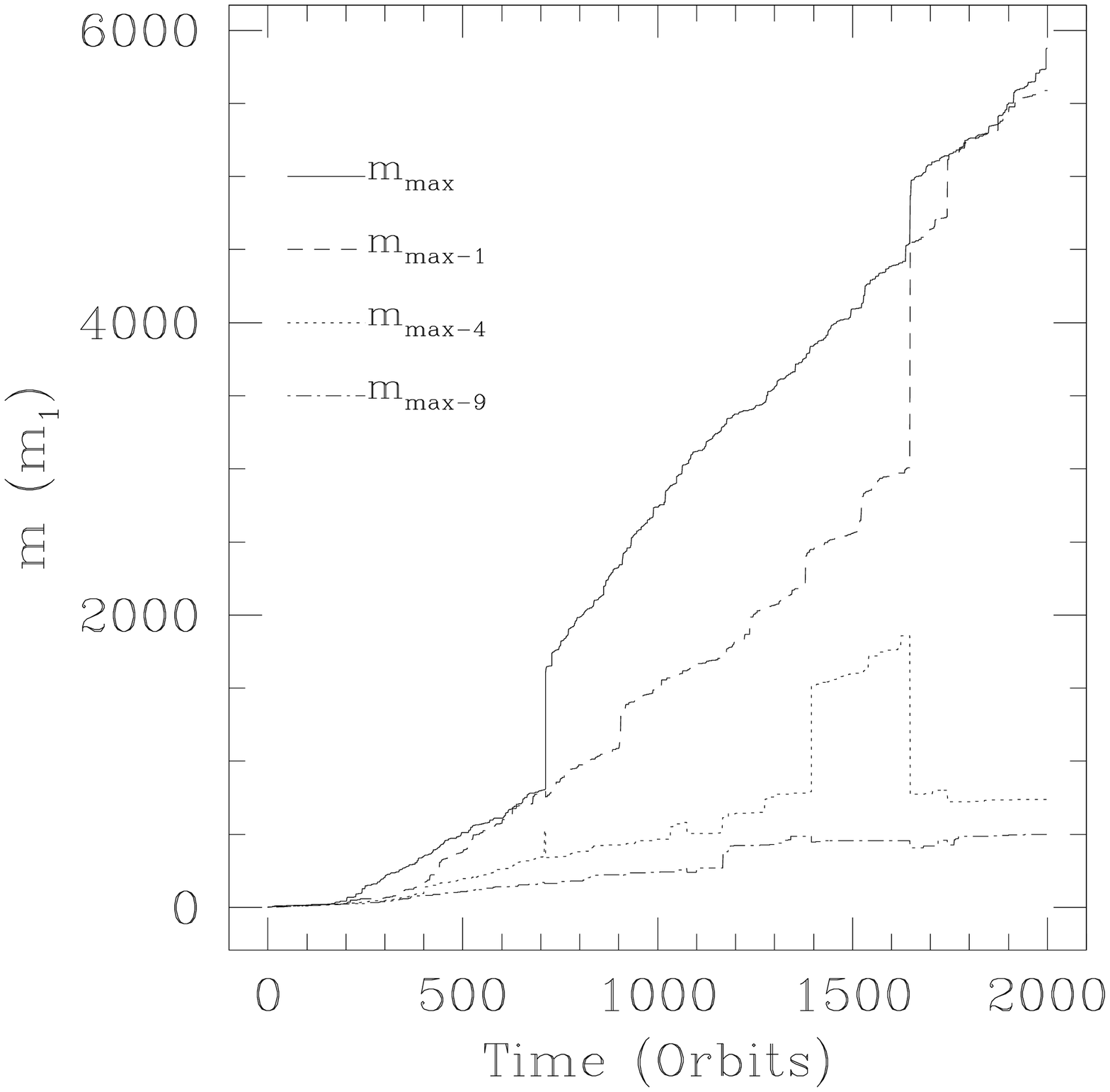}
\figcaption[]{\label{fig:app-lm} \small{Evolution of some of the most massive particles in the L$_1$ run. Note the merger of two $\sim 800 m_1$ objects at 713 orbits. The three largest particle end up with masses in excess of 5000 $m_1$.}}
\medskip

In Fig.\ \ref{fig:app-vrms} we show the evolution of $v_{RMS}$
compared to the escape speed of the largest and typical particle in
Simulation L$_1$. After 500 orbits, $v_{RMS}$ appears to grow
linearly, while $v_{esc}^{max}$ begins to level off (except for the
major merger event at 716 orbits). These qualitatively different
growth rates suggest that the larger particles are beginning to
significantly heat the patch. Note that even this dynamical heating is probably
an underestimate, as particles outside the patch with larger mass
should have sheared into this patch by 700 orbits.

\medskip
\epsfxsize=8truecm
\epsfbox{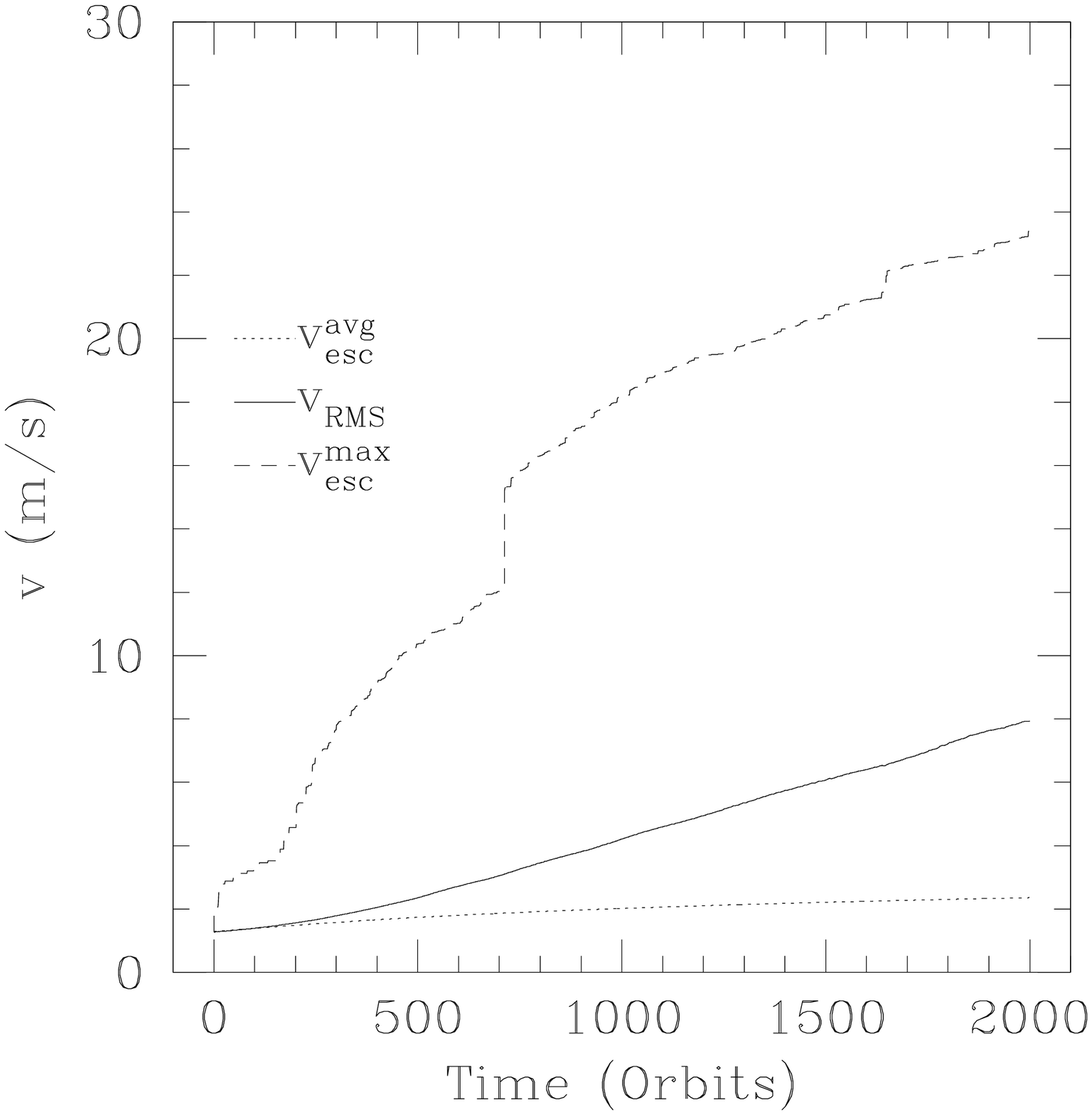}
\figcaption[]{\label{fig:app-vrms} \small{Evolution of the escape speed of the largest particle (dashed line), RMS velocity (solid line), and escape speed of the average mass particle (dotted line) for Simulation L$_1$.}}
\medskip

We compare RMS velocity and mass growth for the three baseline runs in
Fig.\ \ref{fig:app-growthcomp}. The RMS velocities remain close to each
other up to $\sim t_{1/2}$ (top panel), but subsequently diverge. This
divergence corresponds with the masses of the largest particles in
L$_1$ and M$_1$ reaching 100 $m_1$. Note that Simulation M$_1$ has the
highest velocity dispersion from 615 to 1939 orbits. This feature occurs
because the largest particle in M$_1$ is nearly as large as that in
L$_1$ (middle panel), but since the M$_1$ patch is smaller than L$_1$,
the largest particle in M$_1$ contains a larger fraction of the patch
mass (bottom panel), and is therefore a more effective stirrer (see
below). Although the largest particle in Simulation S$_1$ contains
approximately the same fraction of the total mass as the particles in
the other patches, its actual mass is considerably smaller, and,
hence, the velocity dispersion in S$_1$ remains lower than the others.

\medskip
\epsfxsize=8truecm
\epsfbox{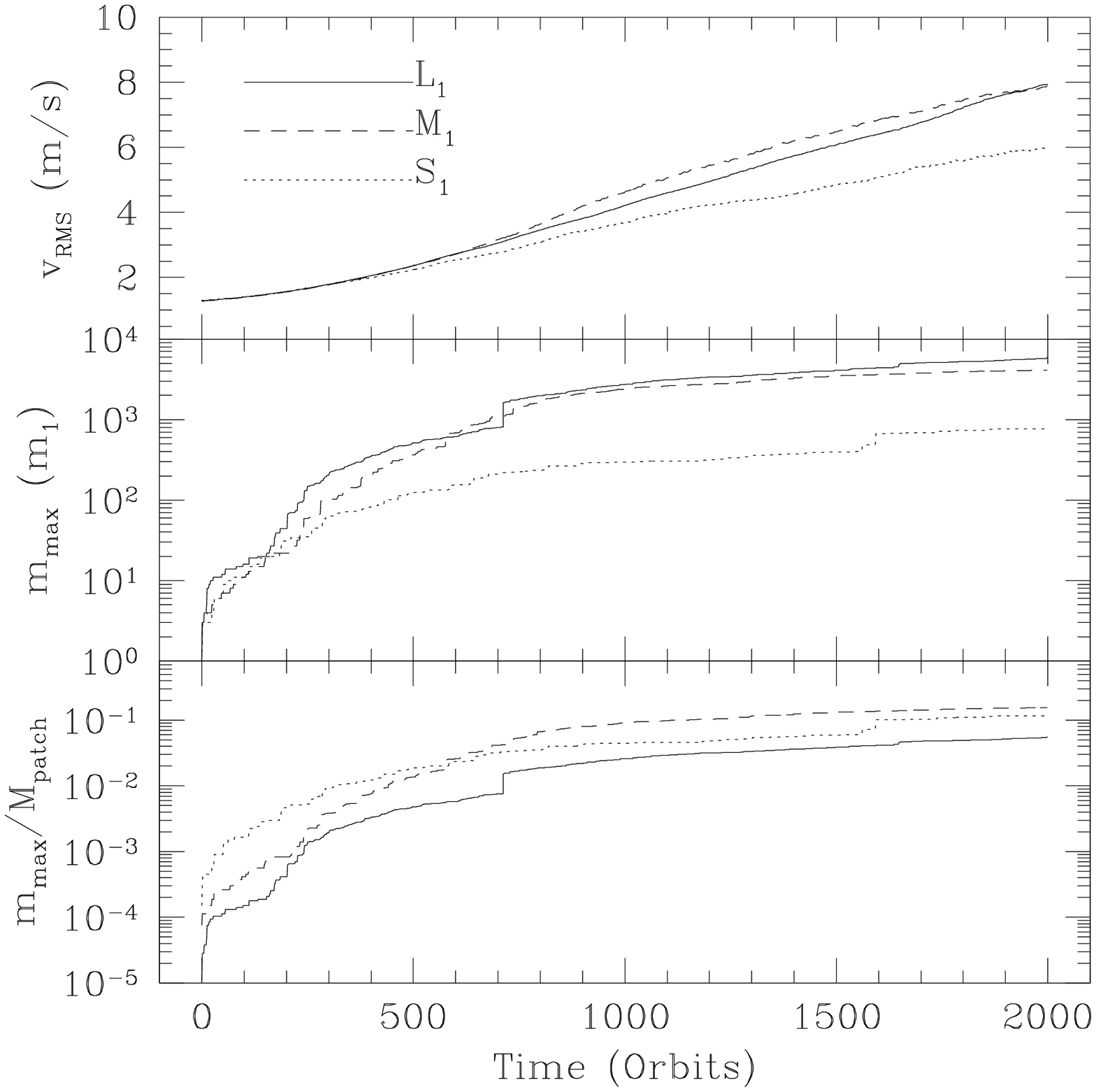}
\figcaption[]{\label{fig:app-growthcomp} \small{\textit{Top:} Evolution of the RMS velocities for the three baseline models. The values for Simulations L$_1$ and M$_1$ stay relatively close to each other, but the S$_1$ value remains lower. \textit{Middle:} Comparison of the growths of the largest particles in the baseline simulations. The largest particles in L$_1$ and M$_1$ remain within about a factor of 2 each other, while the largest particle in S$_1$ lags by about an order of magnitude after 1000 orbits. \textit{Bottom:} Fraction of the patch mass absorbed into the largest particle. At all times, the fractions remain within a factor of a few of each other.}}
\medskip

In Fig.\ \ref{fig:app-F_g} we see the evolution of the three
gravitational focusing factors in the baseline simulations. All appear
to grow, reach a maximum, and then decrease. The peak is at about 300
orbits for S$_1$ and 600 for M$_1$. The largest run L$_1$ is double-peaked, at 400 orbits and 700 orbits. The turnover at
400 orbits in L$_1$ may be a statistical fluke that is dramatically
corrected at 716 orbits, or it may be that the turnover at 400 orbits
is real, and that the event at 716 orbits is anomalous. The curves in
Fig.\ \ref{fig:app-lm} suggest the former. At $t=200$ the largest
particle in L$_1$ begins to grow significantly faster than the second
largest, even though the two have approximately the same mass. Then at
$t = 400$, the difference between the two begins to decrease
suddenly. The two are almost identical at $t = 700$ when two large
particle merge at 716 orbits. At this point, the largest particle once
again becomes significantly larger.

\medskip
\epsfxsize=8truecm
\epsfbox{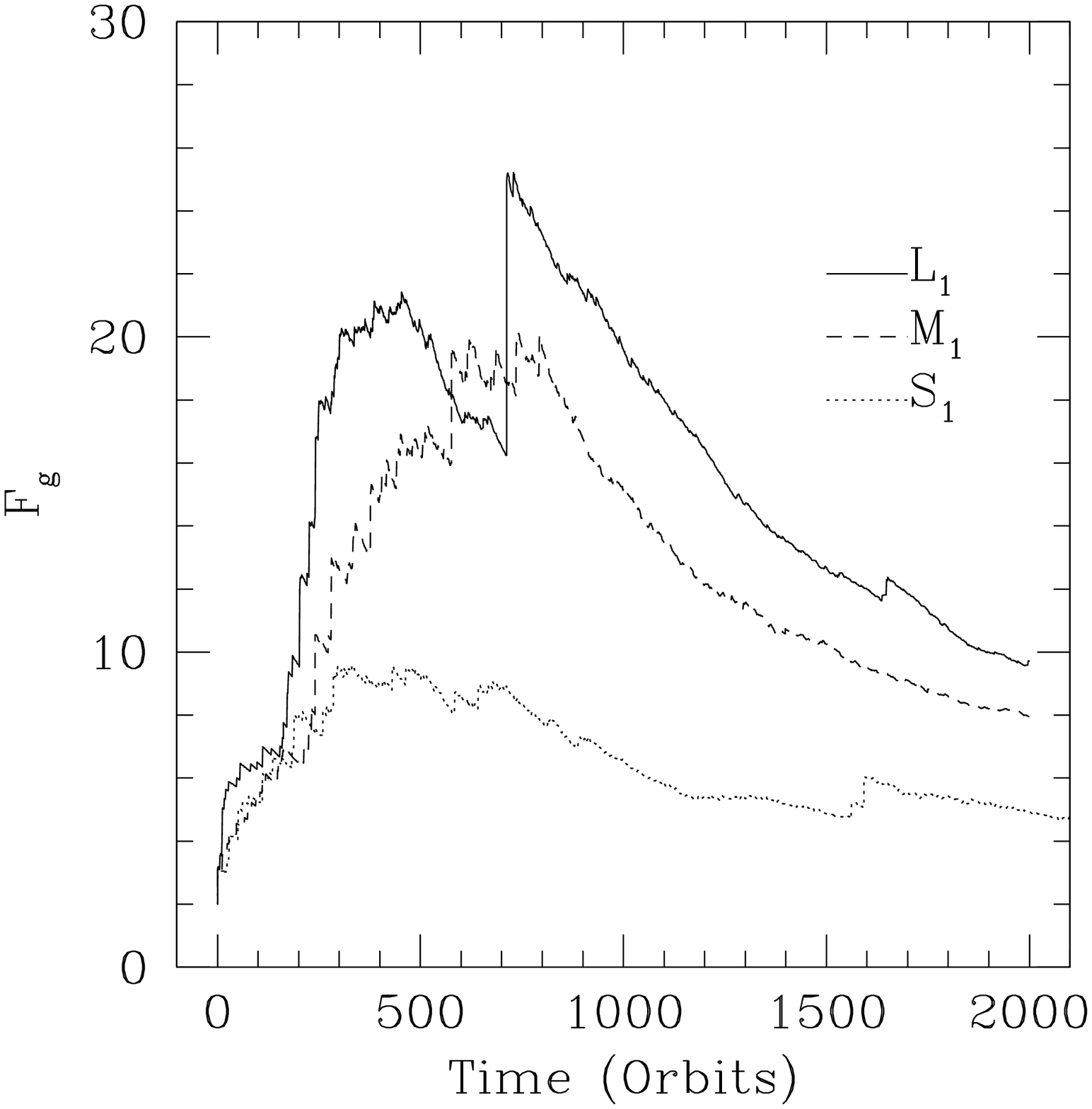}
\figcaption[]{\label{fig:app-F_g} \small{Evolution of $F_g$ as a function of time for the three baseline simulations.}}
\medskip

In Fig.\ \ref{fig:app-eiratio} we show the evolution of $e_{RMS}/\sin
i_{RMS}$. After initially dropping from 2.35 (see Fig.\
\ref{fig:eq-eiratio}), the value in L$_1$ remains close to 2.1 for the
duration of the simulation, suggesting the initial drop is a transient
effect. However, in M$_1$ and S$_1$, the evolution is steady from 200
to 600 or 900 orbits, respectively, and then appears to drop
monotonically at later times. The overall drop represents about a 10\%
change.

\medskip
\epsfxsize=8truecm
\epsfbox{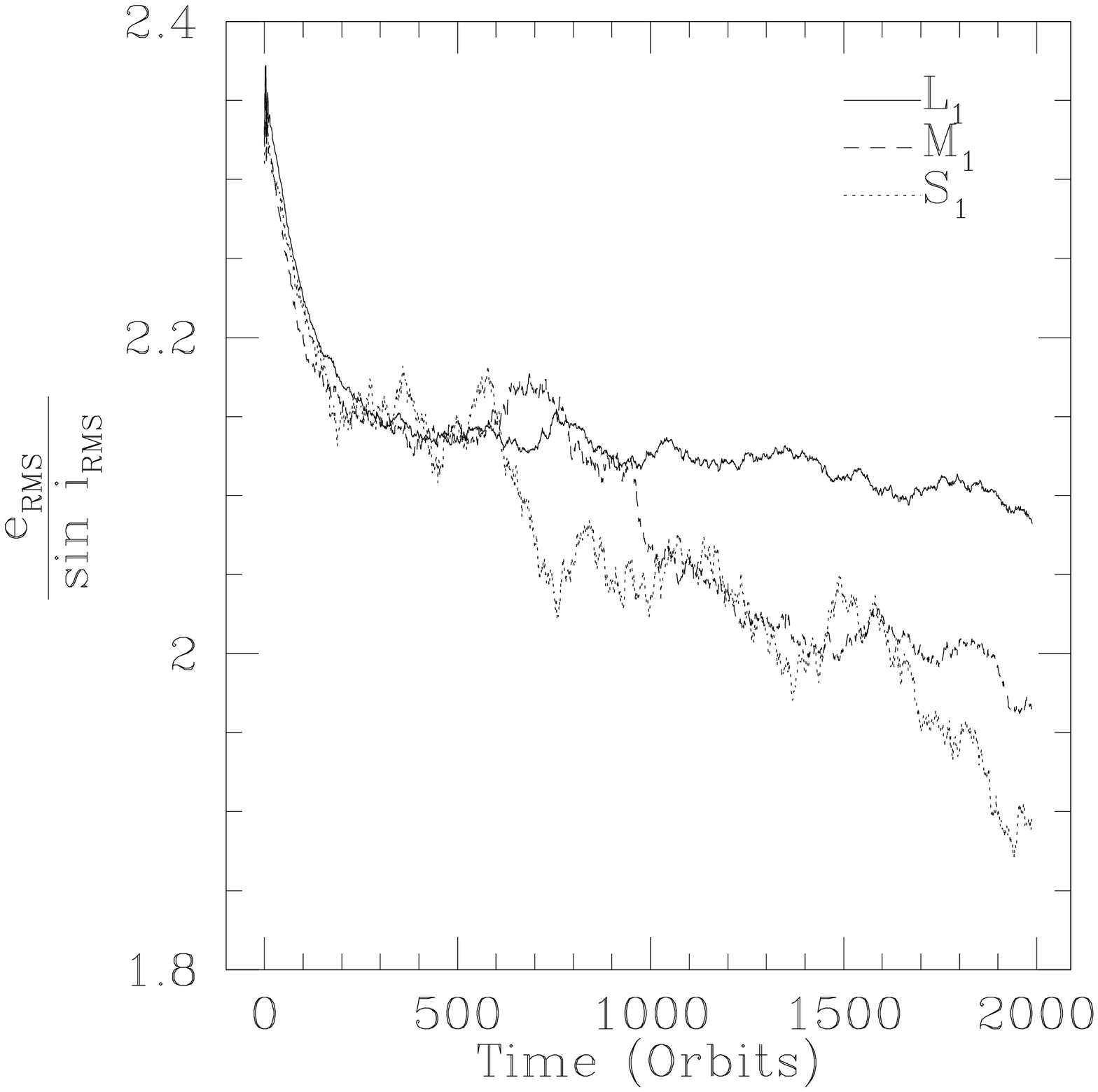}
\figcaption[]{\label{fig:app-eiratio} \small{The ratio of $e_{RMS}$ to $\sin i_{RMS}$ as
a function of time. The ratio in L$_1$ remains slightly over 2 for the duration
of the simulation, but the ratio for M$_1$ and S$_1$ trends down.}}
\medskip

In Fig.\ \ref{fig:app-dynfric} we examine dynamical friction in the
baseline runs at 2000 orbits. We saw in Fig.\ \ref{fig:eq-dynfric} that there was a general trend of decreasing velocity with increasing 
mass, although the slope of this trend is so shallow that kinetic 
energies trend higher with increasing mass.  The same general 
pattern is seen at 2000 orbits for large masses ($m \gsim 100m_1$), but 
velocity is independent of mass for smaller masses. Note as well that the mean values for the
small-mass particles are larger at 2000 orbits than at $t_{1/2}$.

\medskip
\epsfxsize=8truecm
\epsfbox{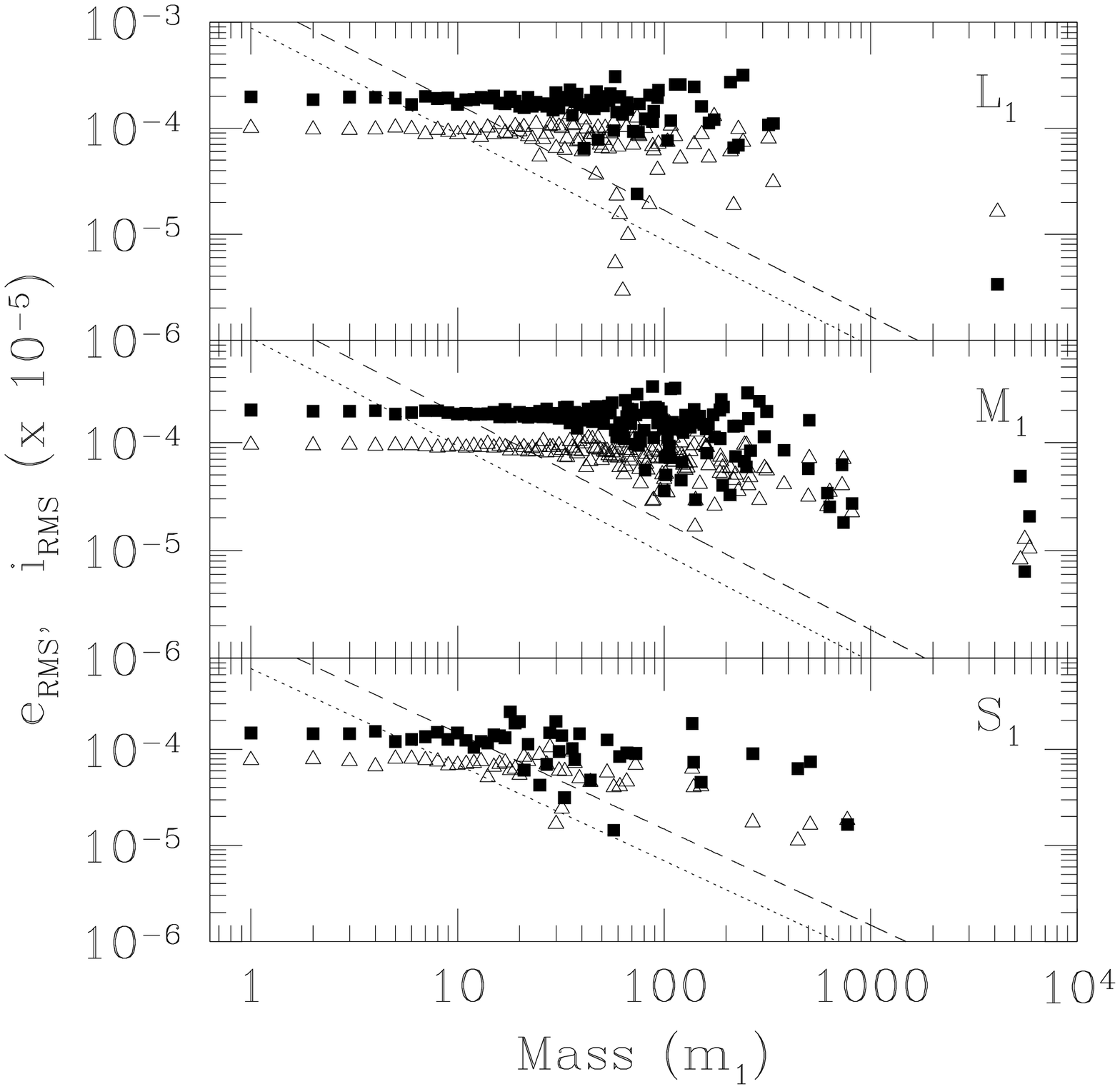}
\figcaption{\label{fig:app-dynfric} \small{The values of $e_{RMS}$ and 
$i_{RMS}$ as a function of mass at 2000 orbits for the three baseline 
simulations. Filled squares are eccentricity, open triangles are 
inclination. For reference, the dashed line represents equipartition of 
energy in $e$, dotted in $i$, normalized to values at $k = 10$.}} \medskip

Next we examine the validity of the L$_1$ patch from $t_{1/2}$ -- 2000
orbits. First we plot the evolution of $u$ and $w$ in Fig.\
\ref{fig:app-uw}. Both values remain small until the major merger
event at 713 orbits. At that time, the values grow significantly,
finally reaching $u = -7$ cm s$^{-1}$, about two orders of magnitude larger
than its value at $t_{1/2}$. This velocity means that the
center-of-mass velocity is roughly 1/70 the total shear
across the patch, and that certainly by the end of the simulation our
assumptions have broken down.

\medskip
\epsfxsize=8truecm
\epsfbox{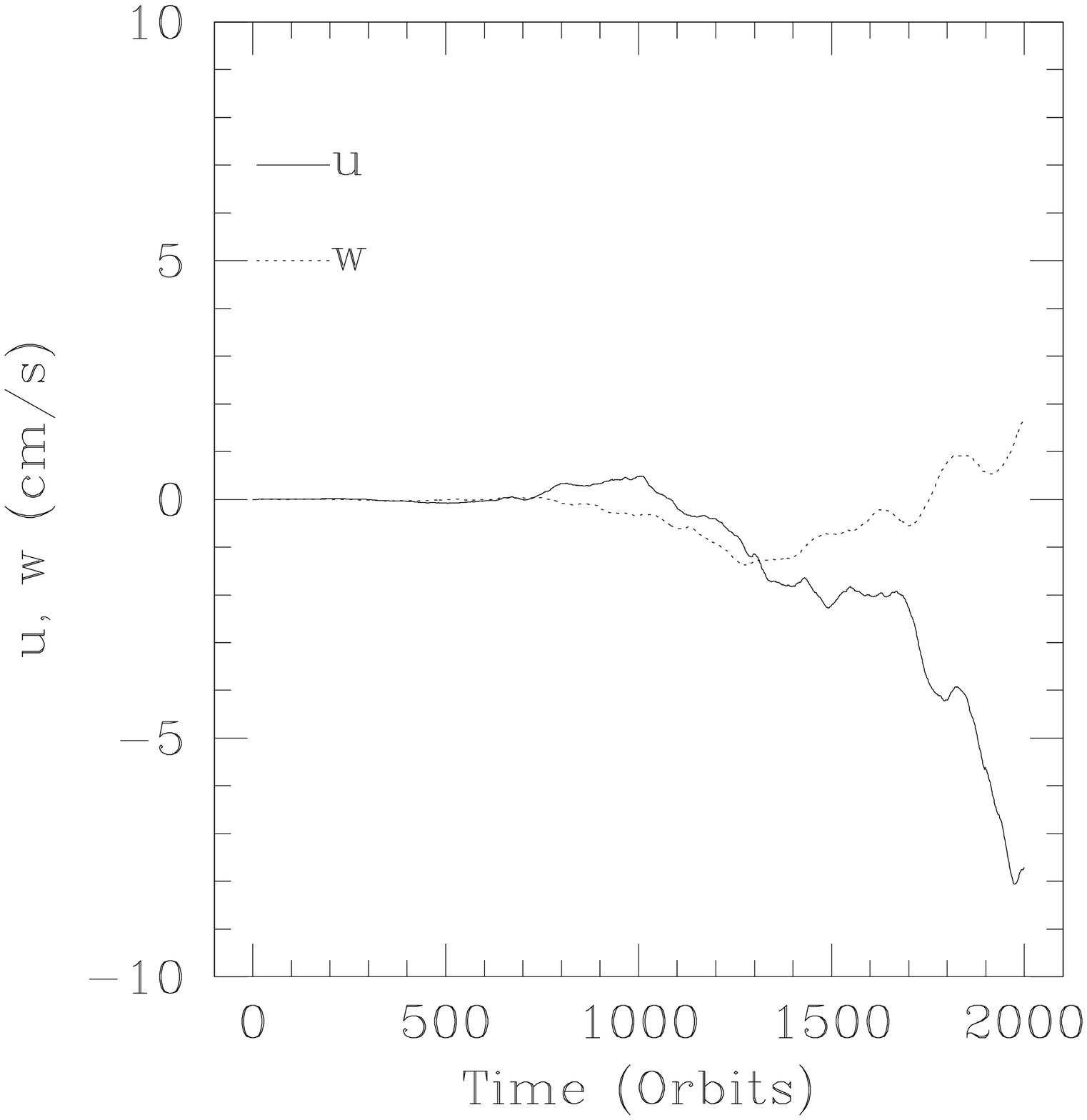}
\figcaption[]{\label{fig:app-uw} \small{Evolution of $u$ and $w$ in Simulation L$_1$. The values grow quickly after the merger at 713 orbits.}}
\medskip

Next we look at how effective different mass bins are at stirring the
patches at different times in Fig.\ \ref{fig:app-m2nll}. We saw in Fig.\
\ref{fig:eq-stir} that by $t_{1/2}$ the distribution of $m^2N_{k}$
was flat, and in Fig.\ \ref{fig:app-m2nll} we see that the slope is
positive at all times after $t_{1/2}$ for all simulations. These
positive slopes indicate that the patch is not large enough
since very large particles not in the patch can substantially affect the velocity
distribution in the patch (see $\S$ 4.4.2).

\medskip
\epsfxsize=8truecm
\epsfbox{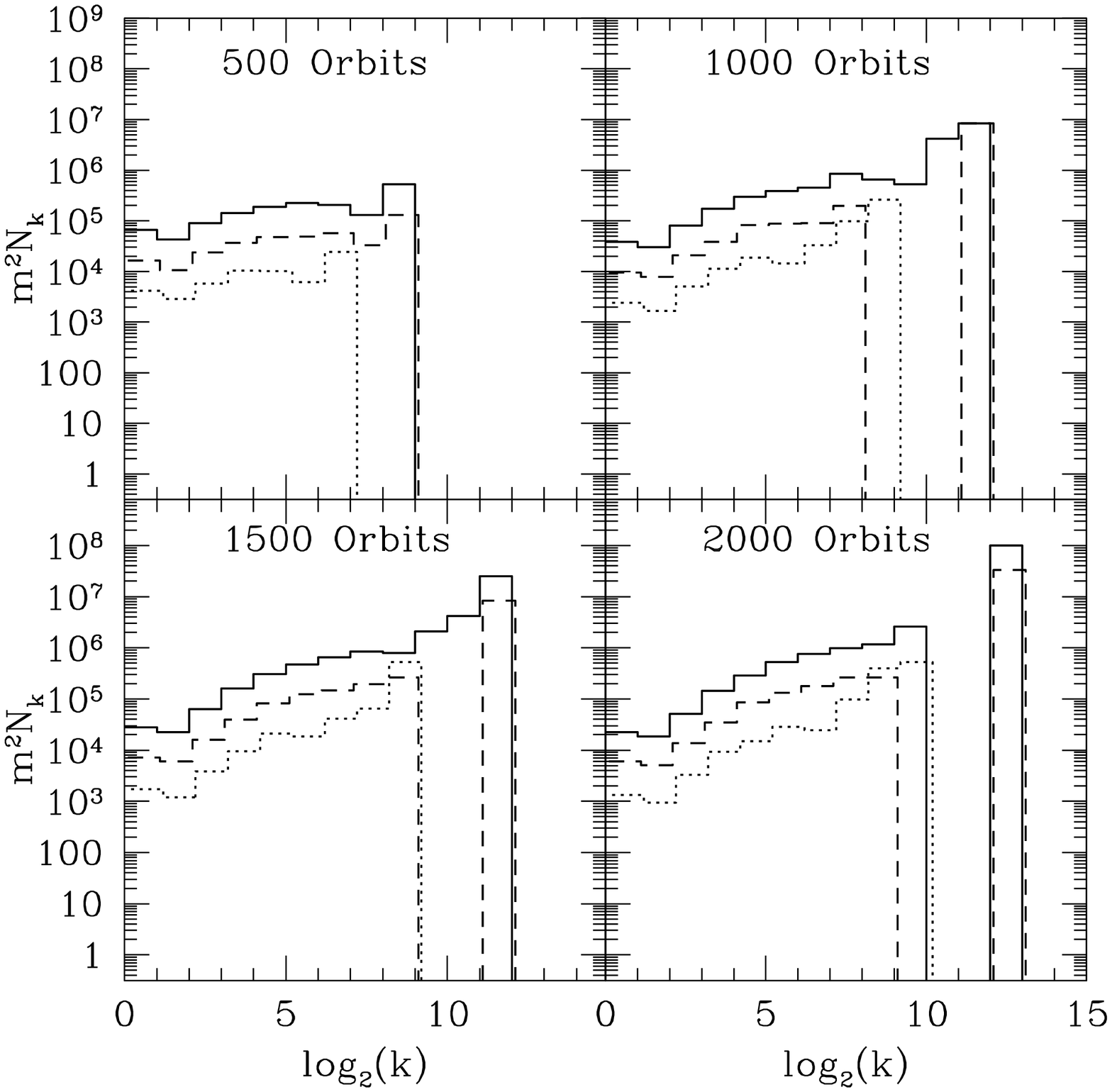}
\figcaption[]{\label{fig:app-m2nll} \small{Distribution of stirring power as a function of mass at four separate times in all baseline simulations. By the end of the simulation, each of the three largest particles in L$_1$ (solid line) are an order of magnitude more effective at stirring the patch than all the other particles combined. Note that the M$_1$ Simulation data (dashed line) are offset by 0.1 and the S$_1$ data (dotted line) by 0.2.}}
\medskip

Finally we examine the value of $S$ in Fig.\
\ref{fig:app-stir}. The value in L$_1$ stays below 0.15 until 713 orbits,
indicating the largest-mass particle is not dominating the stirring in
the patch. However, the merger at 713 orbits creates a particle whose
stirring is about equal to the stirring of the sum of all other
particles in the patch. Therefore at this point, the assumptions of
the patch framework break down, and we can not expect the simulation
to be providing reliable results. As other large particles appear in
L$_1$, $S$ slowly drops. In M$_1$, the value nearly reaches unity at
700 orbits, and then grows quickly to a final value of 17.5. In S$_1$,
$S$ remains below unity for the duration of the simulation, but note
the sudden jump at 1600 orbits.

\medskip
\epsfxsize=8truecm
\epsfbox{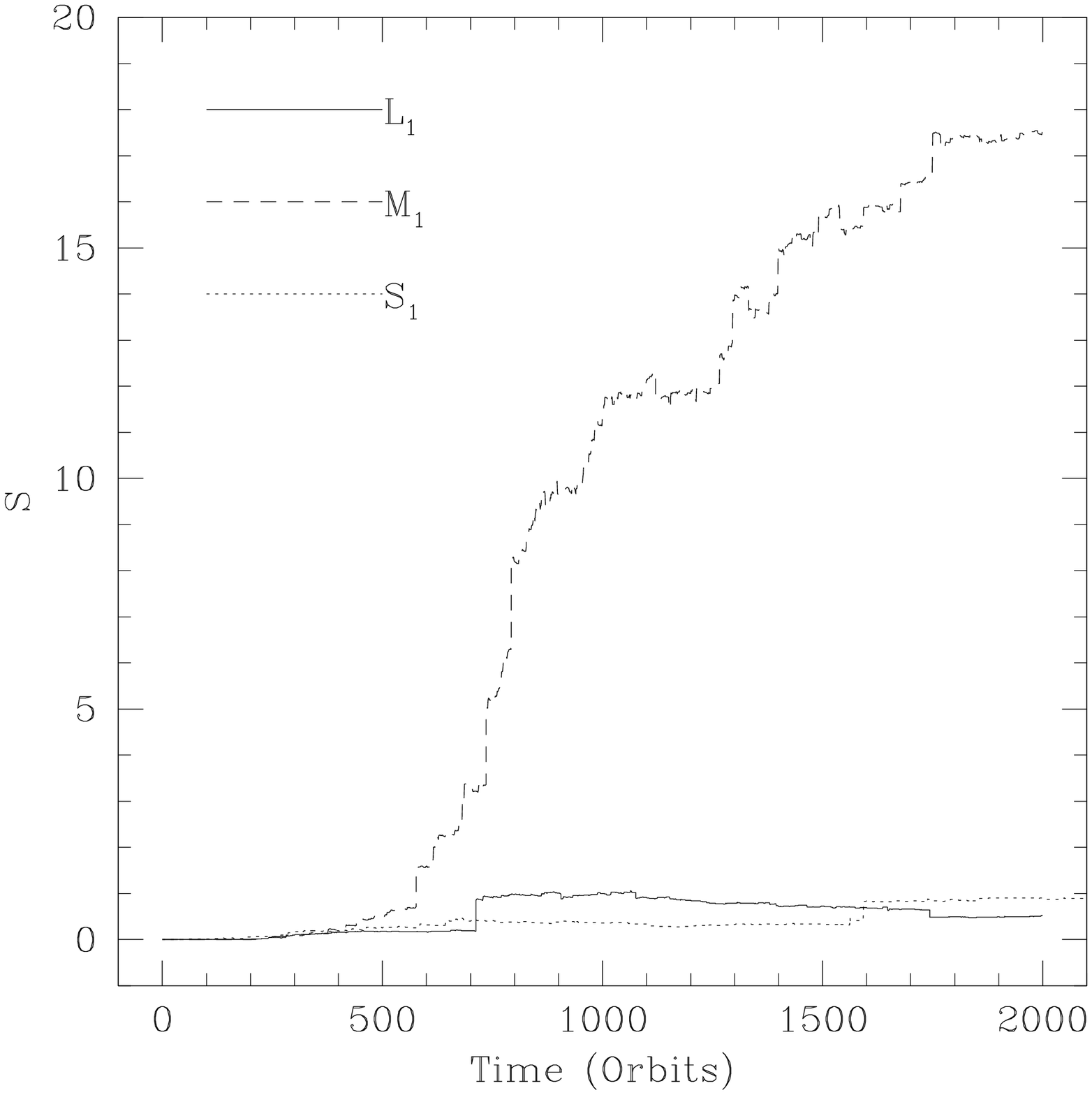}
\figcaption[]{\label{fig:app-stir} \small{Evolution of $S$ as a function of time for the three baseline models. }}
\medskip

The results presented in this appendix reveal the numerous ways in which our patch model breaks down after $t_{1/2}$. The assumptions of small center-of-mass motion, no dominant mass inside the patch, and no outside perturbers have all failed in the three baseline models.

\end{multicols}

%\medskip
\begin{center}
Table 1. Initial Conditions of the Patch Simulations\\
\begin{tabular}{cccccc}
\hline\hline
ID & $N$ & $v_{rms}$ (m s$^{-1}$) & $Z_0$ (km) & $\rho_0$ ($10^{-7}$ g/cm$^3$)\\
\hline
L$_1$ & 106,130 & 1.29 & 475 & 3.1\\
M$_1$ & 26,532 & 1.29 & 475 & 3.1\\
M$_{0.5}$ & 26,532 & 0.65 & 237 & 6.3\\
M$_2$ & 26,532 & 2.6 & 950 & 1.6\\
S$_1$ & 6633 & 1.29 & 475 & 3.1\\ 
\end{tabular}
\end{center}

%\medskip

\begin{center}
Table 2. Results of 1 km Planetesimal Growth at 0.4 AU at $t_{1/2}$\\
\begin{tabular}{cccccc}
\hline\hline
ID & $t_{1/2}$ (orbits) & $m_{max}$ ($m_1$) & $v_{rms}$ (m s$^{-1}$) & $F_g$ & $P_{peak}$ (hr)\\
\hline
L$_1$ & 354 & 276 & 1.93 & 20.2 & 1.05\\
M$_1$ & 349 & 142 & 1.89 & 13.7 & 1.05\\
S$_1$ & 361 & 73 & 1.95 & 9.2 & 1.05\\ 
M$_{0.5}$ & 253 & 202 & 1.86 &  17.8 & 1.65\\
M$_2$ & 542 & 47 & 2.41 & 4.8 & 0.65\\
\end{tabular}
\end{center}

\begin{center}
Table 3. Fit Parameters of the Mass Distributions at $t_{1/2}$\\
\begin{tabular}{ccccccccc}
\hline\hline
ID & $b$ & $\chi_b^2$ & $b'$ & $\chi_{b'}^2$ & $c$ & $\chi_c^2$ & $c'$ & $\chi_{c'}^2$ \\
\hline
L$_1$ & 2.39 & 1693 & 2.62 & 376 & 1.3 & 6224 & 1.9 & 1690\\
M$_1$ & 2.39 & 894 & 2.64 & 168 & 1.2 & 1400 & 1.9 & 853\\
S$_1$ & 2.38 & 492 & 2.65 & 107 & 1.3 & 1073 & 1.9 & 308\\
M$_{0.5}$ & 2.4 & 307 & 2.59 & 74 & 1.2 & 1723 & 2 & 466\\ 
M$_2$ & 2.38 & 489 & 2.61 & 154 & 1.3 & 1440 & 1.7 & 189\\
\end{tabular}
\end{center}

\begin{center}
Table 4. Results of 1 km Planetesimal Growth at 0.4 AU after 2000 orbits\\
\begin{tabular}{cccccc}
\hline\hline
ID & $m_{max}$ ($m_1$) & $v_{rms}$ (m s$^{-1}$) & $F_g$ & $P_{peak}$ (hr)\\
\hline
L$_1$ & 5879 & 7.94 & 9.69 & 0.85\\
M$_1$ & 4117 & 7.89 & 7.95 & 0.75\\
S$_1$ & 773 & 6.00 & 4.95 & 0.85\\ 
\end{tabular}
\end{center}

\begin{center}
Table 5. Fit Parameters of the Mass Distributions at 2000 Orbits\\
\begin{tabular}{ccccccccc}
\hline\hline
ID & $b$ & $\chi_b^2$ & $b'$ & $\chi_{b'}^2$ & $c$ & $\chi_c^2$ & $c'$ & $\chi_{c'}^2$\\
\hline
L$_1$ & 1.88 & 486  & 1.93 & 253 & 8.0 & 11562 & 10.2 & 4511\\
M$_1$ & 1.89 & 141 & 1.96 & 71 & 7.4 & 2783 & 9.8 & 1029\\
S$_1$ & 1.85 & 31 & 1.91 & 20 & 7.9 & 560 & 10.0 & 206\\
\end{tabular}
\end{center}

\end{document}